\documentclass[preprint]{aastex}
\usepackage{lscape} 
\usepackage{amsmath, amsthm,amssymb}
\usepackage{multirow}
\usepackage{url}
\usepackage{natbib}
\usepackage{color}

\newcommand{\full}[0]{{\it full}}
\newcommand{\cold}[0]{{\it cold}}
\newcommand{\hot}[0]{{\it hot}}

\newcommand{\alphaofull}{$\alpha_{\textrm{1}}=1.16_{-0.1}^{+0.17}$}

\newcommand{\alphaohot}{$\alpha_{\textrm{1}}=0.87_{-0.2}^{+0.07}$}
\newcommand{\alphathot}{$\alpha_{\textrm{2}}=0.2_{-0.6}^{+0.1}$}

\newcommand{\Hbhot}{$H_{\textrm{B}}=7.7_{-0.5}^{+1.0}$}

\newcommand{\alphaocold}{$\alpha_{\textrm{1}}=1.5_{-0.2}^{+0.4}$}
\newcommand{\alphatcold}{$\alpha_{\textrm{2}}=0.38_{-0.09}^{+0.05}$}

\newcommand{\Hbcold}{$H_{\textrm{B}}=6.9_{-0.2}^{+0.1}$}

\title{The Absolute Magnitude Distribution of Kuiper Belt Objects}

\author{Wesley C. Fraser {$^{1,2}$}}
\altaffiltext{1}{Herzberg Institute of Astrophysics, 5071 W. Saanich Rd. Victoria, BCV9E 2E7}
\email{wesley.fraser@nrc.ca}
\author{Michael E. Brown {$^2$}}
\altaffiltext{2}{Division of Geological and Planetary Sciences, California Institute of Technology, 1200 E. California Blvd. Pasadena, CA 91125 USA}
\author{Alessandro Morbidelli{$^3$}}
\altaffiltext{3}{Laboratoire Lagrange, UMR7293, UniversitŽ de Nice Sophia-Antipolis, CNRS, Observatoire de la C™te d'Azur.}
\author{Alex Parker{$^4$}}
\altaffiltext{4}{Department of Astronomy, University of California at Berkeley, Berkeley, CA, 94720, USA}
\author{Konstantin Batygin{$^5$}}
\altaffiltext{5}{Institute for Theory and Computation, Harvard-Smithsonian Center for Astrophysics, 60 Garden St. MS 51, Cambridge MA 02138}

\date{} 

\slugcomment{Accepted to ApJ.}
\received{É}
\revised{É}

\begin{abstract}
Here we measure the absolute magnitude distributions (H-distribution) of the dynamically excited and quiescent (hot and cold) Kuiper Belt objects (KBOs), and test if they share the same H-distribution as the Jupiter Trojans. From a compilation of all useable ecliptic surveys, we find that the KBO H-distributions are well described by broken power-laws. The cold population has a bright-end slope, \alphaocold{}, and break magnitude, \Hbcold{} (r'-band). The hot population has a shallower bright-end slope of, \alphaohot{}, and break magnitude \Hbhot{}. Both populations share similar faint end slopes of $\alpha_2\sim0.2$. We estimate the masses of the hot and cold populations are $\sim0.01$ and $\sim3\times10^{-4} \mbox{ M$_{\bigoplus}$}$. The broken power-law fit to the Trojan H-distribution has $\alpha_\textrm{1}=1.0\pm0.2$, $\alpha_\textrm{2}=0.36\pm0.01$, and $H_{\textrm{B}}=8.3$. The KS test reveals that the probability that the Trojans and cold KBOs share the same parent H-distribution is less than 1 in 1000. When the bimodal albedo distribution of the hot objects is accounted for, there is no evidence that the H-distributions of the Trojans and hot KBOs differ. Our findings are in agreement with the predictions of the Nice model in terms of both mass and H-distribution of the hot and Trojan populations. Wide field survey data suggest that the brightest few hot objects, with $H_{\textrm{r'}}\lesssim3$, do not fall on the steep power-law slope of fainter hot objects. Under the standard hierarchical model of planetesimal formation, it is difficult to account for the similar break diameters of the hot and cold populations given the low mass of the cold belt.
\end{abstract}

\begin{document}

\maketitle

\section{Introduction \label{sec:Intro}}
The size distribution is one of the most fundamental properties of a small body population, a property which reflects the collisional processes that have influenced those objects. The size distribution is also one of the most difficult properties to determine primarily as a result of the inability to detect the sizes of most planetesimals directly \citep{Stansberry2008}. As a result, size distributions are usually inferred from the more readily observable apparent magnitude distributions or luminosity functions \citep{Gladman2001}. Such an inference relies on many assumptions about the observed population which when incorrect, can introduce substantial bias into the result \citep{Fraser2008}. Inference from the luminosity function however, can still provide critical insights into the accretion and collisional disruption histories of the observed populations \citep[see for instance][]{Petit2008}.

The luminosity function of the Kuiper belt has been thoroughly studied for more than a decade \citep[for a review, see][]{Petit2008}. These observational efforts have revealed many unexpected properties about the Kuiper belt and its dynamical history. The Kuiper belt exhibits a steep luminosity function that is well represented by a power law for bright objects. That is,  the number of objects brighter than some magnitude, $m$ per square degree on the ecliptic is given by $\Sigma(m)=10^{\alpha\left(m-m_o\right)}$ with the slope $\alpha\sim0.75$ and normalization constant given by $m_o\sim23.4$ in r' \citep{Gladman2001,Fraser2008}. This steep luminosity function slope can be translated to the slope $q=5\alpha+1$ of the underlying size distribution if the size distribution obeys the form $\frac{dn}{dr}\propto r^{-q}$. For the Kuiper belt, this suggests $q\sim4.5$. Such a steep slope may be indicative of a short lived period of accretion lasting on the order of 100 Myr before being halted by mass loss due to the dynamical influence of the gas giant planets \citep{Morbidelli2008}. 

It was \citet{Bernstein2004} who first demonstrated that the power law which describes the luminosity of the bright Kuiper Belt objects (KBOs) does not hold at all sizes, but rather it, and hence the underlying size distribution, breaks to a much shallower slope. The break magnitude, $r'\sim24.5-25$ found by \citet{Fuentes2008,Fraser2009} corresponds to a diameter of $D\sim50-100$~km (assuming 6\% albedos). This break in the size distribution has been interpreted as a remnant of post-accretion collisional disruption which was able to disrupt the majority of objects as large as $\sim100$~km before mass loss froze out the size distribution, halting further accretion \citep{Bernstein2004,Fraser2009c}. It may also be a feature of the primordial size distribution \citep{CampoBagatin2012}.

Insights from the luminosity function are not limited to just the Kuiper Belt as a whole, but rather can be compared to that of other populations to infer relative differences in the collisional histories of the compared populations. \citet{Bernstein2004} was the first to suggest that the dynamically excited, or {\it hot} KBOs, characterized by their high inclinations and eccentricities, exhibit a different size distribution than do the members of the \cold{} population, or those objects found with low inclinations and eccentricities. This finding was supported by \citet{Fraser2010b,Petit2011} both of which found that the hot population exhibits a much shallower luminosity function than the cold population. One interpretation of this observation is that the hot population achieved a much later stage of accretion than did the cold population, resulting in a shallower size distribution with a higher relative amount of mass in the largest bodies as compared to the cold population. 

The utility of this comparison extends further to other populations outside the Kuiper belt region. A popular dynamical model of Kuiper belt formation is one in which most, or all KBOs formed in a region much closer to the Sun, and via a dynamical instability amongst the gas-giant planets, the primordial KBOs were scattered out to their current locales \citep{Gomes2003,Levison2008}. One consequence of this model is the sudden depletion of all Trojan populations of the gas-giant planets, requiring post-instability capture of the scattered objects to account for the observed population. According to that model, the Trojans are repopulated from the same population of objects which were scattered into the Kuiper belt region \citep{Morbidelli2005,Nesvorny2013}. As discussed by \citet{Morbidelli2009},  the size distributions of both the Trojans and the KBOs must reflect the common population from which they both originated. \citet{Fraser2010b} suggested that the luminosity function of the hot population was too shallow to be compatible with the steep luminosity function exhibited by the  Jupiter Trojans, and they concluded that these two populations must not have shared a common primordial predecessor as \citet{Morbidelli2009} suggested. Furthermore, \citet{Fraser2010b} found that the cold population of KBOs exhibited a similar luminosity function as the Trojans. The possibility that the cold population and Trojans share the same precursor populations is extremely difficult to rectify with any known dynamical model suggesting that the similarity in the size distributions of these two populations is merely a coincidence.

One key uncertainty still remains with the findings of \citet{Fraser2010b} and indeed just about all other discussions of the KBO size distribution to date - their reliance on inference from the {\it apparent} luminosity function. Such inference inherently assumes some functional form for the underlying radial, size, and albedo distributions, assumptions which are often untested and even incorrect. This makes the conclusions of such inferences model dependent. One substantial improvement is possible by the inclusion of distance to the observed objects, if such information has been reliably determined. With accurate and calibrated photometry and measured distances to the observed population, the absolute magnitude distribution of that population can be determined. The absolute magnitude distribution has the advantage of removing the distance dependence - a roughly 2 magnitude effect for KBOs - making the inference of the underlying size distribution only sensitive to the unknown albedos of the observed objects. The disadvantage is that, for many available surveys, distances are not well determined, resulting in a reduction in overall data quality compared to the apparent magnitudes alone.

Here we present the first multi-survey, direct determination of the Kuiper Belt apparent magnitude distribution. We tabulate all ecliptic Kuiper Belt surveys from which reliable, absolute photometry, and distance and inclination of moderate quality are available. In Section 2, we discuss the surveys which meet our selection criteria. In Section 3 we present our technique to debias the observations and reconstruct the absolute magnitude distribution. In addition, we present fits of a broken power-law functional form to the resultant distributions. Finally, in this section we present a statistical comparison of the KBO and Jupiter Trojan absolute magnitude distributions. In Section 4 we discuss the consequences of our findings on the origin of the Kuiper belt and Trojan populations and we present our concluding remarks in Section 5.

\section{Datasets and KBO Populations\label{sec:Datasets}}
The goal of this work is to answer three questions:

\begin{enumerate}
\item What are the absolute magnitude distributions or H-distributions of various dynamically distinct KBO populations? 
\item Are the H-distributions of the dynamically distinct KBO populations different?
\item Are the H-distributions of any Kuiper Belt populations compatible with that of the Jupiter Trojans?
\end{enumerate}

To address these questions, we consider here available surveys from which the absolute magnitude distribution of KBOs can be determined. Accurate determination of an object's absolute magnitude, $H$, requires both its distance and its apparent magnitude to be well measured. To produce a debiased absolute magnitude distribution, a well determined detection efficiency for each survey is required. To facilitate comparison of various KBO populations, we further require ecliptic surveys, such that the observations are sensitive to all orbital inclinations, and that each source's inclination can be determined with some certainty. This places strong constraint on which surveys can be used for this work. The survey selection criteria adopted in this work are:

\begin{itemize}
\item on ecliptic survey; observations of ecliptic latitudes $<2^o$ 
\item calibrated detection efficiency with quoted efficiency function
\item calibrated photometry
\item accurately determined source distance and inclination at observation
\end{itemize}

\noindent
The last item is critical; the uncertainty in a source's absolute magnitude as a result of heliocentric distance uncertainty $\delta r$ is given by $\delta H \sim \frac{10}{ln 10}\frac{\delta r}{r}$. As a result, even small distance uncertainties can easily dominate over the typical $\sim0.2$ mag photometric uncertainty in the precision of the absolute magnitudes. In addition, this is also the hardest constraint to quantify. Uncertainty in the determination of a target's orbital parameters depends on the target's observed arclength and the quality of its astrometry, the latter of which is often undetermined in a survey. Other factors such as detection of parallactic motion and quality of the photometry can further influence the errors. As a result, no  hard constraints are easily defined. Rather, we adopt the generic requirement that for ground based surveys, the majority of sources must have arc lengths longer than 24 hours to be considered. Unfortunately, this requirement can still result in inclination errors larger than a few degrees and distance error of a 1-2 tenths of an AU. These uncertainties are still enormous. But we found that this provided the sweet spot between being too constraining and having very little data of use, or being too weak in our constraint and adding too much data of poor quality to our sample. In Section 3 we present a technique for handling these uncertainties in a Monte-Carlo fashion, providing some mitigation against the use of unideal data. 

The surveys which meet our constraints are \citet{Gladman1998, Allen2001,Trujillo2001b,Fuentes2008,Petit2011}. In addition to these surveys, we also consider the surveys of \citet{Bernstein2004} and \citet{Fuentes2009} both of which utilize the detectable parallax motion of KBOs as viewed by the Hubble Space Telescope to provide accurate orbital determination regardless of the short observational arcs.  A problem however, was found for the \citet{Fuentes2009} data; the absolute magnitude distribution that arose from their archival data search was formally incompatible with that found for the amalgamated dataset. This cannot be said about any of the other datasets we considered. Given the nature of the survey and the large distance of a few of the detected sources, it is possible that \citet{Fuentes2009} have stumbled across a new population of distant objects not observed by other surveys. This possibility, however intriguing, seems unlikely and we attribute their result to unreliable distance determinations: we exclude their survey from further consideration. The total areal coverage of all surveys we consider is 294.4 square degrees.

Finally, where possible, when calculating a source's observed absolute magnitude, we utilize the observed colours and absolute magnitudes that are available through the MBOSS database\footnote{\url{http://www.eso.org/~ohainaut/MBOSS/}} and the orbital elements available through the Minor Planet Center\footnote{\url{http://www.minorplanetcenter.net/iau/mpc.html}} (MPC). When these data are available for a given object, that object's orbital parameters and colours as observed by the survey in which it is detected are replaced by the values and uncertainties extracted from the databases.

To address the first and second of our questions, we consider two dynamically distinct Kuiper Belt populations. We first consider the full sample of all detected objects. To avoid any potential biases due to reduced detection efficiency of fast moving objects, we restricted consideration to objects with heliocentric distances $r>29$~AU; we designate this as the \full{} sample. The primary division into two subsamples that we consider, is an inclination division into the dynamically \cold{} and \hot{} objects. We adopt a similar definition as \citet{Bernstein2004}; the \cold{} sample is taken as all objects with inclinations, $i\leq5^\circ$ and observed heliocentric distances approximately bounded by the 3:2 and 2:1 mean motion resonances with Neptune, $38\leq r\leq 48$~AU. The \hot{} sample is the remainder of the \full{} sample not belonging to the \cold{} sample. That is, those objects with inclinations $i>5^o$ and heliocentric distances $38\leq r\leq 48$~AU or objects with all inclinations and distances $29\leq r\leq 38$~AU or $48\leq r$~AU.

As we will discuss below, the division we propose does not completely separate the true hot and cold dynamical classes as those classes overlap in inclination. Such a division will allow mutual contamination amongst the different samples, as dynamically excited objects will have drifted to low inclinations and vice-versa. As such, any differences found between the two populations should be considered as lower limits to the true differences between the two populations.

For the Jupiter Trojan absolute magnitudes, we adopt the absolute magnitudes reported by the MPC. These are standard V(1,0,0) magnitudes. That is, absolute magnitudes in V-band, observed at zero phase-angle. Much of the Trojan population has been surveyed by \citet{Szabo2007}. Later observations however, have demonstrated that the known Trojan sample is incomplete. Comparison of the known sample and that presented by \citet{Szabo2007} reveal no new detections of Trojans with $H\lesssim11$ suggesting the Trojan population is complete (or nearly so) for magnitudes brighter than this.

\section{The Absolute Magnitude Distributions}

In this section, we consider the absolute magnitude distributions of the KBO and Trojan populations. We first present in Section~\ref{sec:debiasing} a method to extract the observed debiased absolute magnitude distribution. In Section~\ref{sec:fitting} we present a maximum likelihood method to fit a model H-distribution to each of the observed populations.

\subsection{Debiasing the Absolute Magnitude Distribution \label{sec:debiasing}}
Here we present histograms of the debiased absolute magnitude distributions. This is for visual presentation purposes only. Modelling of the absolute magnitude distribution will be presented in Section~\ref{sec:fitting}.

For a single, well calibrated survey, determination of the debiased absolute magnitude distribution is simply found by producing a histogram of observed absolute magnitudes, corrected for the effective observing efficiency of the observed absolute magnitude. This can be described as follows. The number of detected objects per square degree with absolute magnitudes $H$ to $H+dH$ and heliocentric distances $r$ to $r+dr$ corresponding to apparent magnitudes $m$ to $m+dm$ is given by

\begin{equation}
n(r,H)dr dH = \Gamma(r)\Sigma(H)\eta(m) dr dH
\label{eq:nrh}
\end{equation}

\noindent
where $\Sigma(H)$ is the intrinsic absolute magnitude distribution, $\Gamma(r)$ is the intrinsic radial distribution, and $\eta(m)$ is a function which describes the effective areal coverage at apparent magnitude $m$. $\eta(m)$ depends on the observing efficiency $\eta_{\textrm{k}(m)}$ and areal coverage $\Omega_{\textrm{k}}$ of each survey. For an individual survey $k$, we adopt the quoted efficiency function of that survey. This usually has the common functional form $\eta_\textrm{k}(m)=\frac{A}{2} \tanh\left(\frac{m-m_{50}}{g}\right)$ where $m_{50}$ is the magnitude at which the detection efficiency is half the peak efficiency, $A$, and the parameter $g$ represents how steeply the efficiency drops from peak to zero. Some surveys however, have a detection efficiency that requires a second parameter $g_2$ in which case the functional form becomes $\eta_\textrm{k}(m)=\frac{A}{4} \tanh\left(\frac{m-m_{50}}{g_1}\right)\tanh\left(\frac{m-m_{50}}{g_2}\right)$. Then the effective areal coverage as a function of magnitude when combining multiple surveys is then just the sum of individual survey detection efficiencies times the areal coverage of that survey. That is

\begin{equation}
\eta(m)=\sum_k \Omega_\textrm{k} \eta_\textrm{k}.
\end{equation}

From Equation~\ref{eq:nrh} it can be seen that, assuming $H$ and $r$ are not correlated, the absolute magnitude distribution is found from the integral of Equation~\ref{eq:nrh} over all $r$. As all datasets we consider were observed near opposition,  we adopt as a good approximation of the absolute magnitude of an object $j$ at distance $r_\textrm{j}$ with apparent magnitude $m_\textrm{j}$ as $H_\textrm{j}=m_\textrm{j}-5\log\left(r_\textrm{j}(r_\textrm{j}-1)\right)$. Thus, the contribution of that object to the observed absolute magnitude distribution is given by

\begin{equation}
\Sigma(H)=\sum_\textrm{j} \left( \int\Gamma(r) \eta\left(H_\textrm{j}+5\log\left(r(r-1)\right)\right) dr \right)^{-1}.
\label{eq:contH}
\end{equation}

\noindent
In a similar fashion, the radial distribution is given by

\begin{equation}
\Gamma(r)=\sum_\textrm{j} \left( \int\Sigma(H) \eta\left(H+5\log\left(r_\textrm{j}(r_\textrm{j}-1)\right)\right) dH \right)^{-1}.
\label{eq:contr}
\end{equation}

It is clear from Equation~\ref{eq:contH} that producing the unbiased absolute magnitude distribution requires knowledge of the underlying radial distribution. Extracting the radial distribution from the dataset we consider is not a simple procedure which is further complicated by the fact that for many of the objects we consider, their distances are only moderately well constrained. Rather, we first adopt a model radial distribution and reconsider the problem of simultaneously extracting both the radial and absolute magnitude distributions in Section~\ref{sec:fitting}. 

As a first attempt at determining $\Sigma(H)$ we evaluate the radial distribution from the  Kuiper belt model produced by the Canada-France Ecliptic Plane Survey \citep[CFEPS - ][]{Petit2011}. As many KBOs are found in mean motion resonances with Neptune, their ecliptic heliocentric distance distribution is not uniform, but rather varies with ecliptic longitude. We assume this variation is symmetric with longitude away from Neptune. We then calculate the radial distribution that each survey would find by binning all objects with longitude from Neptune within 10$^o$ of the survey's longitude from Neptune. As no simple functional form for $\Gamma(r)$ can easily be found, we choose to represent the radial distribution with an interpolated histogram. That is, a radial distribution histogram is produced with bin widths of 2 AU for each of the KBO populations in question. Linear interpolation between bin values is then used to evaluate $\Gamma(r)$ at all distances. With this procedure, we are able to construct estimates of $\Gamma(r)$, one for each observed field, that  accounts for the longitudinal variations of the Kuiper belt. As a practicality, the radial distribution is normalized as $\int \Gamma(r) dr=1$.

When determining the absolute magnitude histogram, we adopt the same latitude and longitude field divisions originally adopted by \citet{Fraser2008} of the surveys presented by \citet{Allen2001} and \citet{Trujillo2001b}. The radial distribution of each field division was then evaluated using the mean longitude of that field. We treat the data presented by \citet{Petit2011} in a similar fashion, and use the internal field designations adopted by the CFEPS survey. As for all other surveys, we adopt the mean longitude of the survey, and evaluate the CFEPS model radial distribution at that point.

Equations~\ref{eq:contH} and \ref{eq:contr} appear simple enough to evaluate. The disparate datasets we consider here present   photometry in different filters and magnitude systems, all of which need to be converted to a common filter; we adopt r'. Thus, for each detected source, a colour conversion to r' must be applied. We use the MBOSS colour database or colour measurements reported by the surveys themselves to convert the apparent magnitude of specific targets to r' when those data are available. Otherwise, the average KBO colours presented in \citet{Fraser2008} are used. This introduces a small $\sim0.1$ magnitude uncertainty in the final r' apparent magnitudes, if colour data for the object is not otherwise available. When dividing the observed sample into dynamical subclasses, the uncertainty in the observed heliocentric distances and inclinations of objects introduce additional sources of error. 

In practice,  when generating the nominal, {\it unbiased} KBO absolute magnitude histograms, uncertainties in the observed parameters $m$, $i$, and $r$ need to be considered.  In an attempt to handle these sources of uncertainty including the colour uncertainty mentioned above, we utilize a Monte Carlo approach. Each source's apparent magnitude $m_\textrm{j}$, heliocentric distance $r_\textrm{j}$, inclination $i_\textrm{j}$, and colour correction are randomly drawn from the uncertainty range of each parameter. For the apparent magnitude, a gaussian distribution is used. For all other parameters, a uniform distribution is used. Each source's absolute r' magnitude is then calculated and its contribution to the absolute magnitude distribution is found from Equation~\ref{eq:contH} and a histogram of the normalized, debiased, differential absolute magnitude distribution is produced. We note that if the randomized inclination or distance places the object outside of the limits of the KBO population in question, then the object is no longer considered as a member of the population for that particular iteration. The procedure is repeated 200 times to determine the 1-$\sigma$ deviation in the histogram values. The resultant differential and cumulative histograms generated with the CFEPS synthetic radial distribution of the \hot{} and \cold{} populations are presented in Figure~\ref{fig:hPlot_CFEPS}. Note: we do not consider heliocentric and geocentric distances separately when generating the H-distribution as the uncertainties in the former are of sufficient size that the additional level of complication does not change the results.

Due to our relatively lax survey constraints, the observed absolute magnitudes can be up to $\sim0.4$ mags uncertain. As a result, to avoid correlated uncertainties on the resultant differential H-distribution histograms, large bin widths are required. We found bin widths of less than 1 magnitude produced correlated uncertainties. It may be that such wide bins may hide or wash-out some underlying structure in the resultant absolute magnitude distributions. The large bin widths are a  reflection of the accuracy with which we can extract the absolute magnitude distribution from the data we consider. 

As the Trojan sample over the absolute magnitude range of interest is complete (or nearly so), the debiased absolute magnitude distribution is simply the observed distribution. The Trojan H-distribution is shown alongside the KBO H-distributions in Figure~\ref{fig:hPlot_CFEPS}.

Despite the numerous uncertainties associated with the data and the coarse nature of the histograms, a few things are immediately apparent from the histograms alone. First and foremost, the \hot{}, \cold{}, and Trojan populations all exhibit the well known break in their magnitude distributions \citep{Bernstein2004, Fuentes2008,Yoshida2008,Fraser2009}. Interestingly, the KBO populations exhibit breaks at absolute magnitudes, $H_{\textrm{r'}}\sim7$,  while the Trojans seem to exhibit a break about a magnitude fainter, $H_{\textrm{r'}}\sim8$. In addition, all three populations are similarly sloped faintward of the break. Brightward of the break, the \cold{} population seems to stand out from the rest with a H-distribution that is steeper than the others. Lastly, it appears that for $H_{\textrm{r'}}\gtrsim7$, both the \hot{} and \cold{} populations as we define them have nearly identical on-ecliptic densities. This is not a result of histogram scaling but a real property of the observations, and is easily seen by the close match of the cumulative histograms faintward of $H\sim 7$.

\subsection{Fitting Procedure \label{sec:fitting}}
To address the first of our questions and quantify the shape  of the absolute magnitude distributions of the hot and cold populations, we utilize a maximum likelihood fit to the observed data. As a reasonable approximation of the debiased differential H-distribution presented in Figure~\ref{fig:hPlot_CFEPS}, we adopt  a broken power-law of the form

\begin{align}
\Sigma(H) & =  10^{\alpha_\textrm{1} \left(H-H_o\right)} \mbox{, for $H<H_\textrm{B}$} \nonumber \\
		& =  10^{\alpha_\textrm{2} \left(H-H_o\right) + \left(\alpha_1-\alpha_2\right)\left(H_\textrm{B}-H_\textrm{o}\right)} \mbox{, for $H>H_\textrm{B}$} 
		\label{eq:Hdist}
\end{align}

\noindent
where $\alpha_\textrm{1}$ and $\alpha_\textrm{2}$ are the power-law slopes for objects brighter and fainter than the transition, or {\it break} magnitude $H_\textrm{B}$, and $H_o$ is a normalization constant. In the approximation that object size does not correlate with distance, an approximation that appears to zeroth order to be true \citep{Stansberry2008}, then the slopes at the bright and faint ends of the observed apparent magnitude distribution are the same as that of the absolute magnitude distribution, $\alpha_1$ and $\alpha_2$. We apply the fits in two ways. For the first, we adopt the synthetic radial distribution extracted from the CFEPS Kuiper belt model discussed above, which at least approximately considers the longitudinal variations in the radial distribution. For this fit, our only free parameters are those of the absolute magnitude distribution. That is, $\alpha_{\textrm{1}},\alpha_{\textrm{2}},H_{\textrm{o}}$, and $H_{\textrm{B}}$.

We preform a second series of fits in which we fit an average radial distribution to the observed data. That is, we fit a single radial distribution as an average of that observed for all fields. With this we will be able to evaluate the importance of the accuracy with which we know the radial distribution as well as how significant longitudinal variations in the radial distribution are when determining the absolute magnitude distribution. We parameterize the fitted average radial distribution in the same way we parameterized the synthetic radial distribution; the radial distribution takes specific values $\Gamma\left(r_{\textrm{c}}\right)=\Gamma_\textrm{c}$ at specific distances $r_\textrm{c}$ and linear interpolation is used to evaluate at distances between the distances $r_\textrm{c}$. Again, the radial distribution is normalized as $\int \Gamma(r|\Gamma_\textrm{1},...,\Gamma_\textrm{c},...) dr =1$ such that the amplitude of the observed distributions is governed only by the parameter $H_\textrm{o}$. Our adopted form provides a potentially more realistic representation of the observed radial distribution than a discrete histogram while avoiding any assumptions about the underlying distribution, other than to assume that the distribution is approximately continuous at distances $r>29$~AU. The distances at which $r_\textrm{c}$ are set are chosen such that for a uniform distribution of objects, the number of objects contained in the  conical volume between each distance is equal. Alternative options were explored, and no qualitative differences were found. For this second set of fits, our free parameters are then the luminosity function parameters $\alpha_{\textrm{1}},\alpha_{\textrm{2}}, H_{\textrm{o}}$, and $H_{\textrm{B}}$, as well as the radial parameters $\Gamma_\textrm{c}$ - five for the \cold{} and \hot{} samples, and six for the \full{}.

A maximum likelihood technique was used to fit Equation~\ref{eq:Hdist} to the observed absolute magnitude distributions. We adopt the same functional form for the likelihood as that presented by \citet{Loredo2004} who gives a complete derivation of the functional form. In the Bayesian framework, the likelihood that the H-distribution parameters $(\alpha_1,\alpha_2,H_o,H_B)$ and for the second set of fits, the set of radial distribution parameters $\Gamma_\textrm{c}$, will produce the observed distribution of objects from surveys $k=(1,..., n)$ is given by

\begin{equation}
L(\alpha_1,\alpha_2,H_o,H_B,\gamma_\textrm{1},...\gamma_\textrm{c},...) = \sum_{k=1}^{n} e^{-\tilde{N}_\textrm{k}} \prod_{j=1}^{N_k} P\left(H_{\textrm{j,k}}r_{\textrm{j,k}}|\alpha_1,\alpha_2,H_o,H_B,\Gamma_\textrm{1},...\Gamma_\textrm{c},...\right).
\label{eq:likelihood}
\end{equation}

\noindent
$N_\textrm{k}$ is the number of objects observed in survey $k$ while $\tilde{N}_\textrm{k}$ is the number of objects expected to be observed by that survey given a particular set of H and radial distribution parameters, and is given by 

\begin{equation}
\tilde{N}_k=\Omega_\textrm{k} \int \eta_\textrm{k}(m) \int \Sigma(H|\alpha_1,\alpha_2,H_o,H_B) \Gamma(r|\Gamma_\textrm{1},...\Gamma_\textrm{c}..) dr dm
\end{equation}

\noindent
where $\Omega_\textrm{k}$ is the areal coverage of survey $k$, $\eta_k(m)$ is the detection efficiency in survey $k$ of an object with apparent magnitude $m$ - we use magnitude in r'-band - which, at opposition is approximately given by $m=H+5 log\left(r(r-1)\right)$.

In Equation~\ref{eq:likelihood}, $P\left(H_{\textrm{j,k}},r_{\textrm{j,k}}|\alpha_1,\alpha_2,H_o,H_B,\Gamma_\textrm{1},...\Gamma_\textrm{c},...\right)$ is the probability of detecting an object with a given  absolute magnitude $H_\textrm{j}$ and distance $r_{\textrm{j}}$ of survey $k$ {\it given} that the object has already been observed. To derive an expression for this, consider the probability of detecting an object at distance $r$, and absolute magnitude $H$. The probability can be written as $P(H,r)=P(H|r)P(r)$, the probability of observing $H$ given a distance $r$ times the probability of observing an object at that distance. If $H$ and $r$ were uncorrelated, then we could write $P(H|r)=\Sigma(H)$ and $P(r)=\Gamma(r)$. But recall that the observable is $m$ not $H$. Further, $m$ and $r$ are imperfect measurements. Thus, we need to write $P(H|r)$ as a function of $m$ and integrate over $m$ and $r$. Substituting $H=m-5\log\left(r(r-1)\right)$, we find the final expression is

\begin{equation}
P\left(m_{\textrm{j,k}},r_{\textrm{j,k}}|\alpha_1,\alpha_2,H_o,H_B,\Gamma_\textrm{1},...\Gamma_\textrm{c},...\right)=\int \int \Sigma\left(m'-5\log\left(r'(r'-1)\right)|\alpha_1,\alpha_2,H_o,H_B\right) \epsilon_{\textrm{j,k}}(m') \Gamma\left(r'|\Gamma_\textrm{1},...\Gamma_\textrm{c},...\right) \gamma_{\textrm{j,k}}(r') dm' dr'.
\label{eq:P}
\end{equation}

where $\Sigma(H|\alpha_1,\alpha_2,H_o,H_B)$ is given by Equation~\ref{eq:Hdist}. $\epsilon_{\textrm{j,k}}(H)$ and $\gamma_{\textrm{j,k}}(r)$ are functional representations of the uncertainty in the observed magnitude $m_{\textrm{j,k}}$ and heliocentric distance $r_{\textrm{j,k}}$ of object $j$ from survey $k$. We adopt gaussian representations for both. Further, we adopt uniform priors on the H-distribution parameters. Finally, for the second set of fits, when the radial distribution was fitted along with the H-distribution we require that the $\Gamma_{c}$ parameters do not take negative values.

To evaluate the quality of Equation~\ref{eq:Hdist} as a representation of the observed H-distributions, we utilize the maximum likelihood value of the fit, $L_{obs}$, and Monte-Carlo simulations. From the best-fit $H$ and $r$ distribution parameters, for each survey we randomly sample a number of objects equal to that observed, consistent with the efficiency parameters of that survey. This random sample is then fit with our maximum likelihood technique, and the maximum likelihood value, $L_{\textrm{ran}}$ of the fit to the random data is recorded. This is repeated to generate a distribution of likelihood values given the best-fit parameters of the observed sample, and from this, the probability $P(L_{\textrm{ran}}>L_\textrm{obs})$ of finding a random maximum likelihood greater than the observed value is found. Values of $P(L_\textrm{ran}>L_\textrm{obs})$ near 0 or 1 indicate a poor fit. 

It should be pointed out that an acceptable alternative approach to determine $P(L_{\textrm{ran}}>L_\textrm{obs})$ would make use of bootstrapping in place of the Monte-Carlo sampling we adopt. Bootstrapping may be considered preferable because it makes use of real detections. No significant differences between the two approaches were found. The exception is that the Monte-Carlo approach seems to produce a broader range of $L_{\textrm{ran}}$ values, and as such, likely provides a slightly more robust test result than bootstrapping, probably as a result of a moderately small dataset.

Uncertainties on the fitted parameters were generated using a Markov-Chain Maximum Likelihood routine. The {\it emcee} software package \citep{Foreman-Mackey2013} was used to produce a large sample of $H$ and $r$ distribution points distributed according to the likelihoods of the \cold{}, \hot{}, and \full{} samples given by Equation~\ref{eq:likelihood}. Then for each parameter, histograms of the posterior distribution marginalized over all other parameters were produced. The uncertainties on each parameter were then adopted as the upper and lower limits which contain 67\% of the marginalized posterior likelihood space, with equal areas outside those limits. {Uncertainties computed from the fits which adopt the radial distribution extracted from the CFEPS model do not fairly reflect the uncertainty in the radial distribution itself. As such, for those fits, we adopt the parameter uncertainties of the fits which treat $\Gamma_\textrm{c}$ as free parameters.

\subsection{Fit Results \label{sec:fitresults}}
Here we present the results of our maximum likelihood fits to the observed H-distributions. In the following sections, we will discuss the fits to the \full{}, \hot{} and \cold{} samples, inference of the H-distribution at larger sizes, and compare these fits to that of the Jupiter Trojans.

\subsubsection{The \hot{}, \cold{}, and \full{} Samples}
Differential and cumulative H and radial distribution histograms for the \cold{}, and \hot{} samples along with the best-fit functions are presented in Figures~\ref{fig:hPlot_CFEPS}, \ref{fig:hPlot} and \ref{fig:rPlot}. The values and uncertainties of the best-fit parameters of the H distribution are presented in Table~\ref{tab:fits}. The marginalized posterior likelihood distributions of those parameters are shown in and Figures~\ref{fig:intervalsCold} and \ref{fig:intervalsHot}.

Examination of Figure~\ref{fig:hPlot} reveals that all three KBO samples exhibit obvious breaks in their absolute magnitude distributions. The inability for a single power-law to describe the distributions of the \full{}, \hot{}, and \cold{} populations is confirmed by the values of $P(L_{\textrm{ran}}>L_\textrm{obs})$ of the best-fit power-laws, which have values less than 1 in 100 for the \full{} and \cold{} populations, and 0.02 for the \hot{} population. We find that for the \cold{} and \hot{} populations, the best-fit broken power-laws of Equation~\ref{eq:Hdist} are adequate descriptions of the observations (see Table~\ref{tab:fits}). For the \full{} sample, the broken power-law provides only a moderately acceptable fit to the observations, with $P(L_{\textrm{ran}}>L_\textrm{obs})=0.16$.

The best-fit broken power-law of the \hot{} population has a large object slope \alphaohot{} that breaks to a slope \alphathot{} at magnitude \Hbhot{}. The best-fit broken power-law to the \cold{} sample, has large object slope \alphaocold{} that breaks at a magnitude \Hbcold{} to a slope similar to the faint-end slope of the \hot{} population, \alphatcold{}. 

The fits which utilize the CFEPS model to evaluate the KBO radial distributions produce similar results as the fits which include a global radial distribution as free parameters (see Table~\ref{tab:fits}). For the cold{} sample, the fits are nearly identical, with best-fit parameter values matching well within the $1-\sigma$ uncertainties. For the \hot{} population, the fits with the CFEPS model result in shallower slopes and a fainter break magnitude. The discrepancy may be a result of the global radial distribution used in one set of fits, which does not account for the longitudinal structure in the Kuiper Belt compared to the CFEPS radial model which includes some longitudinal variations in the model. The discrepancy may also be caused by inaccuracies in the CFEPS model. The discrepancy on the parameters however, is still within the $1-\sigma$ errors on the parameters themselves, and so may just be a result of data quality. What ever the cause, additional data are required to refine the \hot{} population break magnitude.

From the fits to the \cold{} and \hot{} samples, it is clear that both populations exhibit different large object slopes. This is in agreement with \citet{Bernstein2004} who first suggested that the hot and cold KBOs posses different size distributions. The results of our fits suggest that the \full{} population is not very well described by a broken power-law, while the \cold{} and \hot{} populations are. By considering the populations with a separate size distribution, the maximum likelihood value of the fits is improved over that when both are treated simultaneously with a single H-distribution. To determine if the improvement is significant, we turn to the likelihood ratio test. The likelihood ratio is ratio of likelihoods of the more complex model - \cold{} and \hot{} treated separately - and the simpler model - the \full{} sample, $R_{\textrm{obs}}=\frac{L_{\textrm{hot}}L_{\textrm{cold}}}{L_{\textrm{full}}}$. We performed a series of Monte Carlo simulations in which a random cold and hot sample was drawn from the best-fit radial and H-distribution of the observed \full{} sample. The random samples were then fit independently, and together, and the random likelihood ratio of the samples, $R_{\textrm{sim}}$ was recorded. This process was repeated 100 times to determine the probability $P(R_{\textrm{sim}}>R_{\textrm{obs}})$. That is, the probability that the observed improvement in fit quality was only chance. We found $P(R_{\textrm{sim}}>R_{\textrm{obs}})=14\%$. 

In addition to the likelihood ratio test, we make use of the Kuiper-variant Kolmogorov-Smirnov (KKS) test \citep{Smith2002} to determine if the observed sample is consistent with one population, or two. The KKS value between the observed samples, the best-fit broken power-laws to the \hot{} and \cold{} samples was evaluated as follows. Random objects were drawn from the best-fit broken power-law of the \full{} sample and accepted to a master sample with detection efficiency equal to that of the observed populations. The sampling was repeated until a master sample 2500 times as large as the observed samples was drawn. The KKS value between the master sample and the observed \full{} sample, $V_{\textrm{obs}}$ was found. Then, from the master sample, samples of equal size as the \hot{} and \cold{} samples were drawn, with magnitudes scattered according to the observed absolute magnitude uncertainties, and the random KKS value, $V_{\textrm{ran}}$ between that random sample and the master sample was evaluated. This process was repeated 500 times to evaluate $P(V_{\textrm{ran}}>V_{\textrm{obs}})$, the probability of finding a KKS value as low as observed given that the \hot{} and \cold{} samples actually posses the different, observed H-distributions. We found $P(V_{\textrm{ran}}>V_{\textrm{obs}})=3$\%.

All evidence (the quality of the broken power-law fit to the \full{} sample, the improvement in fit quality when \cold{} and \hot{} are taken separately, and the probability of finding the observed KKS value) supports the idea that the \cold{} and \hot{} samples posses different H-distributions.

To determine what the absolute magnitude distributions are telling us about the underlying size distributions, and to fairly compare the hot and cold populations, one must consider the albedos of the populations in question. We have collected albedo data from \citet{Thomas2000}, \citet{ Grundy2005}, \citet{Stansberry2008}, \citet{Brown2006b}, \citet{Brucker2009}, \citet{Lellouch2010}, \citet{Lim2010}, \citet{Mommert2012}, \citet{Ortiz2012}, \citet{Pal2012}, \citet{Santos-Sans2012}, \citet{Stansberry2012}, \citet{Vilenius2012}, \citet{Bauer2013}, \citet{Braga-Ribas2013}, \citet{Brown2013}, and \cite{Fornasier2013}. In Figures~\ref{fig:alb_vs_H} and \ref{fig:alb_vs_colour} we present visual albedo versus absolute R magnitude and (B-R) colour of \hot{} and \cold{} KBOs. As there is no obvious difference in colour or albedo properties of centaurs and equal sized objects with $r>29$~AU\citep{Fraser2012} we include centaurs with $r>7$~AU for consideration of the \hot{} sample albedos. 

Three previously discovered properties of KBO albedos are immediately apparent from these figures:

\begin{itemize}
\item \cold{} objects exhibit higher albedos than do similar sized \hot{} objects \citep{Brucker2009}.
\item the small \hot{} objects have a bimodal colour distribution, with each colour group exhibiting a unique mean albedo \citep{Stansberry2008,Fraser2012}.
\item \hot{} objects exhibit a trend of decreasing albedo with increasing absolute magnitude  \citep{Stansberry2008}.
\end{itemize}
\noindent

The sample of albedos of \cold{} objects is presented in Figure~\ref{fig:alb_vs_H}. Two objects,  (79360) Sila-Nunam and (119951) 2002 KX14, with $p\sim9\%$ have low albedos compared to the other objects. \citet{Parker2010a} has demonstrated that wide binaries such as Sila-Nunam cannot be objects which have been scattered to their current orbits by the gas giants as presumably most other \hot{} objects have been. Thus, it seems that this object is a genuine \cold{} population member.  As we will discuss, unlike Sila-Nunam, 119951 is most likely a low-$i$ \hot{} object rather than a \cold{} object.

A simple division in inclination to divide the \hot{} and \cold{} samples, as we have adopted here does not correctly separate those populations, but rather produces mixtures of the two underlying populations. Fortunately, the magnitude of mixing can be estimated. To do this, we adopt the inclination distribution of the CFEPS survey. They model the inclination distribution of each of the hot and cold populations with the probability distribution originally presented by \citet{Brown2001}

\begin{equation}
P(i) \propto \sin\left(i\right) e^{-\frac{1}{2}\left(\frac{i}{\sigma}\right)^2}.
\label{eq:incdist}
\end{equation}

\noindent
Here, $\sigma$ is the width of the population in question and we adopt the best-fit values of the \cold{} and \hot{} populations of $\sigma_\textrm{C}=2.6^o$ and $\sigma_\textrm{H}=16^o$ respectively \citep{Petit2011,Gladman2012}. As shown by \citet{Brown2001}, assuming circular orbits, the fraction of time an object with inclination $i$ is found below an ecliptic latitude $\beta$ is

\begin{equation}
F(\beta)=\frac{2}{\pi}\sin^{-1}\left[\min\left(\frac{\sin\beta}{\sin i},1\right)\right].
\label{eq:PI}
\end{equation}

\noindent
The observed fraction of either population found below a given inclination division is then determined by integrating the product of Equations~\ref{eq:incdist} and \ref{eq:PI} from 0 to the value of the division, in our case, $5^o$. For the observed fraction above this, the integration is carried out above the $5^o$ inclination cut. 

We find that below $2^o$ ecliptic latitude, 94\% of the total population below that latitude belongs to the \cold{} population, and 23\% of the total \hot{} population is observed below $5^o$ inclination. Considering the observed H-distributions, objects with $H=4.5$ - like 119951 - are roughly 5 times more likely to be a low-$i$ interloper from the \hot{} population rather than a true \cold{} member. When 119951 is excluded,  the distribution of measured \cold{} albedos is consistent with a constant value.

When albedo is uncorrelated with size, the slope of the absolute magnitude distribution translates directly to the logarithmic slope of the underlying size distribution. That is, the assumption of a power-law size distribution of the form $\frac{dn}{dr}\propto r^{-q}$ results in the absolute magnitude distribution of the form we adopted in Equation~\ref{eq:Hdist}, $\frac{dn}{dH}\propto 10^{\alpha H}$ with a linear relation between the slopes of $q=5 \alpha +1$. Thus, we infer that the underlying size distribution of the \cold{} sample is well represented by a broken power-law with large and small object slopes $q_1=8.2\pm1.5$ and $q_2=2.9\pm0.3$.

Over the range of magnitudes $5\leq H \leq 10$ where we have measured the \cold{} and \hot{} H-distributions, the \cold{} sample has a weighted mean albedo of $15\pm2\%$ (excluding 119951). Ignoring phase effects, the absolute magnitude of an object relates to its diameter, D, and albedo (ratio of reflected to incident light), $p$ as $H=K-2.5\log\left( \left(\frac{D}{100\mbox{ km}}\right)\right)^2p$. From the quoted albedos, diameters, and magnitudes presented by \citet{Vilenius2012}, the median value of the normalization is $K=5.61\pm0.03$. Thus, the break magnitude of the \cold{} sample, \Hbcold{}, corresponds to a break diameter of $D_\textrm{B}=140\pm10$~km.

For the \hot{} population, the trend of decreasing albedo with increasing magnitude which is obvious for objects with $H\lesssim4$ may continue over the range of absolute magnitudes probed by our fits; over the range $4\lesssim H \lesssim 10$ the Spearman rank correlation test reports only a 10\% probability of randomly drawing the observed negative correlation. The effect of this trend  is to decrease the inferred slope of the underlying size distribution by only $\sim3\%$ over that inferred with uniform albedos. This effect is significantly smaller than the uncertainty in the fitted slopes, and is ignored. The inferred broken power-law size distribution has large and small object slopes $q_1=5.3_{-1}^{+0.4}$ and $q_2=2_{-1}^{+0.5}$.

The red and blue, or neutral \hot{} objects, over the same absolute magnitude range, exhibit mean albedos of $12\pm1\%$ and $6.0\pm0.5\%$, respectively. When inferring the underlying break diameter of the \hot{} population, the bimodal albedo distribution of that population must be accounted for. This can be understood by considering the net detection efficiency as a function of absolute magnitude which is presented in Figure~\ref{fig:Heff}. As can be seen, the efficiency drops precipitously with absolute magnitude. At the break diameter of the \hot{} population, due to their higher albedos, the red objects are $\sim3$ times more likely to be included in the \hot{} sample than the neutral objects. Thus, it is more appropriate to consider the mean albedo of the red objects when inferring the underlying break diameter, which for the \hot{} population is $D_\textrm{B}=110_{-80}^{+10}$~km.
  
From these results, a few things about the size distributions of the \hot{} and \cold{} samples are apparent. First, within the precision of the fits, both samples exhibit the same slopes for objects smaller than their breaks. The same cannot be said about the slopes for larger objects; the \cold{} population exhibits a much steeper slope. Finally, within the precision of the fits, both populations exhibit compatible break diameters. Though, due to a lack of observed small objects, the break magnitude and diameter of the \hot{} population is significantly more uncertain, and it could be that the break diameter of the \hot{} sample is actually a factor of $\sim2$ smaller than that of the \cold{} sample.

From the inferred size distributions, we can estimate the mass of the \hot{} and \cold{} Kuiper belt populations. To do this, we integrate over the best-fit H-distributions. We assume albedos of 15\% and 6\% for the \cold{} and \hot{} populations, respectively, and apply corrections due to the inclination distribution derived above using the best-fit CFEPS inclination distribution widths of $2.6^o$ and $16^o$. We also assume that the size distribution faintward of our detection limits is well described by a power-law, and that the mass is bounded. That is, $q<4$. Assuming a material density of $1\mbox{ g cm$^{-3}$}$, we find that the masses of the \cold{} and \hot{} belts are $3\times10^{-4}$ and 0.01 Earth masses, respectively. With our assumptions, the size distribution faintward of our detection limits contributes roughly 50\% uncertainty to the masses. Another important uncertainty  results from the uncertain densities of objects; we attribute a 50\% uncertainty to these values due to object density. We find that the mass of the \hot{} belt is at least a factor of 10, and could be as much as a factor of 100 more than that of the \cold{} belt.

It should be noted that, as the observed \cold{} distribution is actually a mix of the intrinsic \hot{} and \cold{} populations -  the observed sample is $\Sigma_{i<5^o}\left(H\right) = 0.94 \Sigma_{\textrm{cold}}\left(H\right) + 0.23\Sigma_{\textrm{hot}}\left(H\right)$ - and that the observed \cold{} population H-distribution is steeper than that of the \hot{} population, it must be that the intrinsic  \cold{} H-distribution is  even steeper than observed. The inverse must also be true for the intrinsic \hot{} population. For example, we find that when considering the predicted mixing caused by our inclination cuts, we find a satisfactory fit to the {\it observed} \cold{} and \hot{} populations when  large object slopes roughly 5\% steeper and 15\% shallower than the fitted \cold{} and \hot{} values are used. These values are within the 1-$\sigma$ uncertainty range on the fitted parameters, and as such, consideration of mixing due to inclination cuts will only be important once significantly more data become available.

\subsubsection{The Bright End of the \hot{} H-distribution}

As no wide field survey is available that meets our data requirements, the data we present here do not probe the brightest few magnitudes of the KBO H-distributions. Additional constraint on the bright end however, can be found from  \citet{Sheppard2011} and \citet{Rabinowitz2012}. The focus of these surveys were previously un-surveyed regions of the Southern sky. These surveys do not meet our first requirement, that of on ecliptic observations. Under the assumption that there is no significant size distribution-inclination correlation for objects in the hot population, a rough estimate of the on-ecliptic H-distribution can still be found by correcting each objectÕs contribution to the all-sky H-distribution by the probability of finding an object at that latitude. We adopt the hot population inclination distribution discussed above and determine the latitude distribution with the same technique as \citet{Brown2001}. Note: this implicitly assumes that all latitudes which were observed by these surveys were surveyed equally, which is not entirely true at the upper and lower extents of their observations. Further, the observations of \citet{Rabinowitz2012} suffer from two potential weaknesses: their photometry was calibrated with respect to the USNO-B catalog, and their detection efficiency was determined from field asteroids with potentially uncertain magnitudes. As a result, the H-distributions derived from these observations should only be considered approximate.

We apply the same data cuts as applied to the data we discuss in Section~\ref{sec:Datasets}. That is, we only consider objects with heliocentric distances $r\geq 29 $~AU. \citet{Sheppard2011} present a well determined detection efficiency. Thus, for that survey we consider objects whose probability of detection was greater than 50\%. The detection efficiency of \citet{Rabinowitz2012} is uncertain, and as a result, we are forced to restrict use of their data to detections with $R<20.5$ where their efficiency appears approximately constant with magnitude. This ensures that the resultant H-distribution shape is not affected by poorly determined efficiencies, but only its normalization. The results of these cuts are 7 objects from \citet{Sheppard2011} and 11 objects from \citet{Rabinowitz2012}, none of which belong to the \cold{} sample.

Where available, MBOSS colours were used to convert the R-band absolute magnitudes presented by \citet{Sheppard2011} to r'. Otherwise, the average $<r'-R>=0.26$ presented in \citet{Fraser2008} was used. 

As discussed above, the largest KBOs have significantly higher albedos than smaller objects. \citet{Fraser2008} suggested that the albedo-size trend could be described by $\rho \propto D^{\beta}$. As can be seen, for size distribution considerations, the \hot{} object albedos are adequately described by $\rho=\left(\frac{D}{250 \mbox{ km}}\right)^2+6$\%. We can use this relation to correct the observed absolute magnitudes to the effective absolute magnitudes, the values that would be found if all objects had the same albedos.

The effective H-distributions inferred from the observations of \citet{Sheppard2011} and \citet{Rabinowitz2012} is presented in Figure~\ref{fig:hPlot_wSheppard}. As can be seen the effective H-distribution of the \hot{} population is consistent with the best-fit power-law with slope \alphaohot. It is only the two brightest objects in the observations, Pluto and Eris, that deviate away from that power-law; the probability of drawing two Eris-sized objects from the best-fit broken power-law of the \hot{} distribution is only $\sim5\%$. We find similar results if we use $\beta=1.5$ instead. That is, for reasonable choices in $\beta$, brightward of the break, there is no evidence that the effective H-distribution deviates from the best-fit power-law for all but the brightest few objects - those with $H_{\textrm{r',eff}}\lesssim3$ or $H_{\textrm{r'}}\sim0.5$. This result can be restated that there is no evidence for a deviation in the hot KBO size distribution away from a power-law for objects with diameters smaller than $D\sim1000$~km and larger than the break diameter, $D\sim140$~km. We must remind the reader however, that these are only approximate results. Confirmation of the power-law behaviour - and any deviations away from it - will require a survey which is equally complete at all ecliptic latitudes with well calibrated detection efficiency and photometry.

\subsubsection{Why So Steep?}
Other than the estimate from \citet{Petit2011}, previous inferences of the KBO size distribution have been from the observed {\it apparent} magnitude distributions of KBOs \citep[see][for a review]{Petit2008}. These past surveys typically observed power-law slopes of $\alpha<0.6$ for the \hot{} population and $\sim0.8$ for the \cold{} population \citep{Bernstein2004,Fraser2010b}. These values are much shallower than the large object slopes we have inferred from the observed absolute magnitude distributions. Our findings are similar to the results of \citet{Petit2011} who first self consistently analyzed the absolute magnitude distribution and found steeper slopes for their observed on-ecliptic population for $H{\textrm{r'}}<8.5$, a slope  similar to the large object slope we find for the \full{} sample of \alphaofull{}. 

The reason why the absolute magnitude distributions reveal steeper slopes than inferred from the apparent magnitude distributions is simply a result of the large distance over which the Kuiper Belt is distributed. The apparent magnitude of an object at 35 AU will be 1.5 mags brighter than the same object at a distance of 50 AU. This broad distance range over which the bulk of KBOs are found results in a spread of the absolute break-magnitude of $H\sim7$ into  $22.5\lesssim r'\lesssim24$ in apparent magnitude space. This can be seen in Figure~\ref{fig:LF}. The apparent magnitude distributions, which are the convolution of the best-fit $H$ and $r$ distributions provide good descriptions of the observed apparent magnitude distributions, and exhibit broad roll-overs.  Only brighter then $r'\sim22.5$ or fainter than $r'\sim24$ are the true bright and faint object slopes apparent. 

Power-law fits to observations which occupy part of the range $22.5\leq r' \leq24$ (the majority of past works) suffer a perceived flattening of the best-fit slope compared to the actual large object slope. We tested this with some basic Monte Carlo simulations of a typical on-ecliptic luminosity function survey. We adopted the best-fit $H$ and $r$ distributions of the \hot{} sample, and simulated the apparent magnitude distribution that would be observed in the survey presented by \citet{Fraser2010b} which found a very shallow slope of $\alpha=0.35\pm0.2$ for objects with apparent magnitudes $21.5 \leq r \leq 24.5$. We randomly generated a number of objects equal to that observed in their survey and fit a power-law to the randomly generated sample, and repeated this process 1000 times. The slopes found by our simulations had a mean of 0.57 with sample deviation of 0.1. These values are typical of past efforts, though slopes as shallow as that observed by \citet{Fraser2010b} occurred in only 1\% of our simulations. Thus, we conclude that only with prior knowledge of the radial distribution, will analysis of the apparent magnitude distribution reveal the correct slope of the underlying absolute magnitude and size distributions.

\subsubsection{Comparison with Other Recent Works}
\citet{Schwamb2014} have compiled data from wide-field surveys with moderate quality photometry, none of which meet our efficiency or photometric requirements. From those data, they found that the bright end of the \hot{} KBO apparent luminosity function, for objects with $m\lesssim24$ in R-band is consistent with a power-law of slope $\sim0.8$. This is similar to our findings here that the \hot{} object absolute magnitude distribution for $H_{\textrm{r'}}<7$ is consistent with a power-law for all but the brightest two known KBOs.

While their cold and hot samples are defined based on their model Kuiper Belt rather than the inclination cut we adopt here, \citet{Petit2011} find similar results as those we present here. Fitting single power-laws to the hot and cold luminosity functions, they find slopes of $\alpha_{\textrm{hot}}=0.8_{-0.2}^{+0.3}$ and $\alpha_{\textrm{cold}}=1.2_{-0.3}^{+0.2}$. In addition,  they conclude that the hot and cold components cannot have the same size distribution at greater than the 99\% confidence. They arrived at this conclusion by a forward modelling approach, in which they assigned model semi-major axis, eccentricity, inclination, perihelion, and absolute magnitude distributions. From those chosen distributions, a number of simulated observed samples were drawn and compared to the real observed sample. Model parameters were varied until acceptable ranges were found. It is not clear how much their significance on the result that the cold and hot populations share different H-distributions is affected by the, admittedly complex, orbital element distribution modelling. But their general findings corroborate ours. That is, all evidences points to the result that the cold and hot H-distributions posses different slopes for $H\lesssim7$.

\citet{Shankman2013} present evidence that the H-distribution of the scattered disk population exhibits a divot, or sudden decrease in the  density of objects below a certain absolute magnitude. They find that the H-distribution of scattered objects (observed for $H_{\textrm{g'}}<9$ and inferred at fainter magnitudes from the Jupiter Family Comets) is well described by power-law with slope $\alpha=0.8$ to the divot magnitude, $H_{\textrm{g'}}=9$. At the divot magnitude, there is a drop in density by a factor of $\sim5$ and faintward of this magnitude, the distribution continues as a power-law with slope $\alpha<0.5$.

The \hot{} sample we consider here does not uniquely contain scattered disk objects. We examined the \hot{} sample for evidence of a divot nonetheless. Adopting the average KBO (g'-r') colour, 0.65, taken from \citet{Petit2011}, the preferred divot magnitude of \citet{Shankman2013} takes value, $H_{\textrm{r'}}=8.3$ in r'. From Figure~\ref{fig:hPlot}, it is clear that no obvious evidence for a divot exists at this magnitude. In fact, adopting our best-fit large object slope for the \hot{} population, we can formally eliminate divots faintward of $H_{\textrm{r'}}=7.7$ at the 1-$\sigma$ level. For magnitudes faintward of the best-fit break magnitude and brightward of H=7.7, the largest drop in density in the form of a divot can be no more than a factor of 3 (1-$\sigma$ limit). Certainly, adding a divot at {\it any} magnitude does not improve the quality of the fits to the observed \hot{} distribution over that of the broken power-law we adopt above. Thus, we conclude that for the \hot{} population, which is a mix of scattered disk and other excited KBOs, we have no evidence for a divot in the observed H-distribution.

In another recent analysis of the KBO size distribution, \citet{Schlichting2013}, compile observations from past luminosity surveys, and convert these observations into a size distribution. They find that the the size distributions of both the \hot{} and \cold{} samples are well described by power-laws with differential slopes $q=4$. This slope is much shallower than the large object slopes we find for either population. Their  slope was found by converting the apparent magnitude distribution into a size distribution by assuming all observed objects were at the same distance.  As a result, their inferred size distribution suffers from the same effect as other past efforts; incorrect knowledge of the underlying size distribution has resulted in a blurring of the true size distribution, and a much shallower slope than in reality. Further evidence of this effect comes from the fact that their size distribution exhibits  no evidence for the break we find, which should occur at $\sim 90$~km in their plots (they assume 4\% albedos for all KBOs). We conclude that the size distribution they discuss is not an accurate representation of the true KBO size distribution.

\subsubsection{The Jupiter Trojan H-distribution} 

The best-fit  broken power-law to the Trojans, along side the H-distribution histogram is shown in Figure~\ref{fig:hPlot}. This histogram has been scaled to place the KBO and Trojan H-distributions on a similar vertical scale to ease comparison between them. The best-fit has large object slope $\alpha_1=1.0\pm0.2$, similar to the value found by \citet{Jewitt2000}. The best-fit breaks to a slope $\alpha_2=0.36\pm0.01$, compatible with the slope found by \citet{Yoshida2007}. The best-fit break magnitude is $H_\textrm{B}=8.4_{-0.1}^{+0.2}$.

We now turn our attention to comparing the Trojan H-distribution to that of the hot and cold KBO populations. As discussed above, one must consider the albedo distributions of each population to ensure equal size scales for comparison. For Trojans with diameters larger than $D\sim60$~km - the same size scale as our KBO samples - their mean R-band albedo is $4.4\pm0.2$\% \citep{Fernandez2009}, lower than the albedos of both the \cold{} and \hot{} KBOs. Thus, the underlying break diameter of the Trojans is $D=136\pm8$~km, very similar to the break diameters of both KBO populations.  As the Trojans also exhibit similar faint object slopes as the KBO populations, it seems any differences between the Trojan and KBOs must lie with the large object slopes.

To get a quantitative measure of the H-distribution differences, we made use of the non-parametric Kuiper-variant Kolmogorov-Smirnov test (KKS) to test the probability, $P(V_{\textrm{ran}}>V_{\textrm{obs}})$, of the null-hypothesis, that the two samples are drawn from the same parent distribution \citep{Press2002}.  During this test, we took care to consider both the albedo differences and colour differences between the KBO and Trojan samples. This was done as follows. Starting from the observed Trojan V absolute magnitudes, a random magnitude was selected. It was then adjusted to the magnitude it would have in the r' by randomly selecting a (r'-V) colour from the mean colour distribution of the Trojans \citep{Szabo2007}. The magnitude was then corrected for its albedo by drawing a random albedo consistent with the albedo distribution of the KBO population in question. Finally, the detection bias of the KBO sample was applied; the random H-magnitude was accepted to the master biased Trojan sample with probability equal to the net H-magnitude efficiency of the KBO sample. This process was repeated until a master sample of 50,000 random magnitudes was generated. From this random sample, the KKS statistic, $V_{\textrm{obs}}$, between the corrected Trojan and KBO sample in question was calculated. 

To determine the value of $P(V_{\textrm{ran}}>V_{\textrm{obs}})$, simulated KBO samples were generated by bootstrapping samples of objects from the master Trojan sample of size equal to the KBO population in question, and scattering those magnitudes according to the observed H-magnitude errors of that population. From each bootstrapped, scattered sample, a KKS value, $V_{\textrm{ran}}$ was calculated. The process was repeated to produce a distribution of KKS values from which $P(V_{\textrm{ran}}>V_{\textrm{obs}})$ was calculated.

When the \hot{} sample was compared with the Trojans, it is necessary to consider the albedo distributions of the neutral and red \hot{} objects separately. Recall that the red objects are approximately 3 times more likely to be observed than the neutral objects. If a simple mean of the entire \hot{} population were adopted in our test, the bias towards higher albedos would not be correctly accounted for, and the test results would be spurious. We adopted the separate albedo distributions by generating half the random master Trojan sample with the albedo distribution of the neutral objects, and half from the red; Gaussian distributions with mean and widths equal to the sample mean and widths of the observed \hot{} object albedos were utilized. We found that the probability that the \hot{} and Trojan samples share the same parent distribution is 38\%. Put simply, when the albedo distributions are fairly accounted for, there is no detectable difference in the H-distributions of the \hot{} KBOs and the Jupiter Trojans. We note that the same conclusion is drawn for neutral to red mixture fractions of 0.1 to 0.9.

When the \cold{} sample was compared to the Trojan population in this manner, the probability $P(V_{\textrm{ran}}>V_{\textrm{obs}})$ was found to be less than 1 in 1000. That is, there is less than 1 in 1000 probability that the \cold{} and Trojan populations are drawn from the same parent distribution.  This is driven by the much steeper large object slope of the cold distribution (\alphaocold). The fact that the \cold{} sample cannot share the same parent distribution as the Trojans, while the \hot{} population can is in agreement with our assertion that the \hot{} and \cold{} KBO samples do not share the same size distribution. 

\section{Discussion \label{sec:Discussion}}

The results we present here have some significant consequences for our understanding of planetesimal growth, and the origin of the Kuiper Belt and Jupiter Trojan populations. We first turn our attention to the origins of the the planetesimal populations.

The Nice model predicts that the Trojans of Jupiter were captured from an original trans-Neptunian disk, during a phase of orbital instability of the giant planets \citep{Morbidelli2005,Nesvorny2013}. A small fraction of the same disk would survive today in the Kuiper belt \citep{Levison2008}.

Previous works, which incorrectly inferred the KBO size distributions from the apparent luminosity functions have suggested that the \hot{} KBO  size distribution exhibited a large object slope too shallow  to be compatible with that of the Jupiter Trojans \citep[see for instance][]{Fraser2010b}. Further, the \cold{} population was found to exhibit a size distribution that looked very similar to that of the Trojans. This was surprising, because in the Nice model the Trojans should have been captured from the hot population, not the cold one. To solve this paradox, \citet{Morbidelli2009} built a model of the original trans-Neptunian disk that was made of two parts. The inner part had a $H$-distribution like that usually attributed to the hot belt, and the outer part had a distribution like that attributed to the cold belt.  When the giant planets became unstable, part of the outer disk populated the current cold population. The rest of the outer disk mixed with the inner disk and from this mixed population both the Trojans and the hot population got implanted in their current locations. Given that the large object H-distribution of the outer disk was steeper than that of the inner disk, the observed mixed population was dominated by the former population for objects with $H\gtrsim6$. This explained why the Trojans appeared to have a distribution as steep as the cold population, leading to the prediction that the hot population should also exhibit a $H$-distribution that is equally steep for $H\gtrsim 6$.

The results from this paper change the situation considerably. We find no evidence that, over the common observable size range, the Trojans and hot KBOs exhibit different size distributions. In fact, all three measured parameters of the H-distributions, the bright and faint object slopes and the break magnitude (when corrected for albedo differences) are indistinguishable within the errors of the measurements. This is in agreement with the findings of the Nice model, that, barring collisional evolution, the Trojans and hot population should share the same size distribution. 

From the simulations presented by \citet{Levison2008}, they determined that roughly 0.1\% of the primordial population from which the hot KBOs originated was actually deposited into the current Kuiper Belt. Barring any additional loss mechanisms, our findings suggest that the primordial population contained roughly 5 to 20 Earth masses of material. This is not dissimilar to the $25 \mbox{ M$_{\bigoplus}$ }$ required to produce the dynamical instability which inevitably populated the hot Kuiper Belt. 

Our findings strongly favour the general scenario put forth by the Nice model, where both the hot KBO and Trojan populations originated from the same primordial population, and were scattered to their current locales from a region between $\sim15$ and 35 AU. The same cannot be said about the cold KBOs. In the model of \citet{Levison2008}, the
primordial disk was truncated at 34 AU, and the cold KBOs were emplaced from within this limit. Levison et al. were
aware that, to explain the differences of the slopes of the size distributions and  of the colours, the cold KBOs
had to be derived from a different region of the primordial disk than the hot KBOs, but their simulations
mostly failed to do so.  Moreover,
\citet{Parker2010a} have demonstrated that the widely separated binary objects in the
cold population could not have undergone scattering off Neptune, unlike in Levison et al. scenario. Thus,
unless a new transport mechanism is found that does not involve close encounters with Neptune and feeds
the cold population from a region of the primordial disk that does not generate hot KBOs, we are left with the
hypothesis that the cold population is local, i.e. it formed in situ.

\citet{Batygin2011} has shown that a primordial population in the cold belt region can survive the
giant planet instability event which populated the hot belt, and experience only minimal excitation, consistent
with the current dynamically quiescent orbits of the cold objects. The problem with this view is that no mass
from this region (or just a small fraction of it) is lost by dynamical removal. But today, the cold Kuiper belt
contains only  $3\times 10^{-4} M_{\oplus}$. It is hard to envision the formation of objects several hundred kilometres in diameter in such a low mass environment. So, where did the mass go? Collisional grinding is not an option:
it could not have produced order-of-magnitude mass loss, particularly given the small collision velocities of cold
KBOs among themselves \citep[see discussions by][]{Morbidelli2008,Nesvorny2011}. Further, \citet{Parker2012} has demonstrated that the wide binary planetesimals in the cold belt are easily disrupted by even moderate collisional evolution; their presence demonstrates that significant collisional evolution of the cold belt did not occur. A remaining challenge with the in-situ scenario is the outer edge of the cold population observed at ~45 AU, beyond which no cold objects are found.

A rough estimate of  the collisional timescale which governs both accretionary and destructive collisional processes, $t_c$ is that $t_c\propto \frac{\Sigma}{a^{3/2}}$, where $\Sigma$ is the surface density of the planetesimal population and $a$ is its semi  major axis. We can use this relation to consider the formation timescale of objects in size equal to the break diameters of both the hot and cold KBO populations. The growth timescale ratio at different disk locations with semi-major axes $a_1$ and $a_2$ is $\frac{t_\textrm{h}}{t_\textrm{c}}\propto \left(\frac{a_1}{a_2}\right)^{\frac{3}{2}} \left(\frac{\Sigma_{2}}{\Sigma_{1}}\right)$. Recent simulations of the growth of 1000~km bodies from $\sim1$~km sized planetesimals by \citet{Kenyon2012} corroborate this behaviour.

It seems most likely that the cold objects formed at their current locales $\sim 40$~AU, while the hot objects were scattered into place from a region between $\sim15$ and 35 AU. In this region, the mass of planetesimals required to populate the hot and Trojan populations is roughly $20-30\mbox {M$_{\bigoplus}$}$ implying a primordial surface density of $\Sigma_{\textrm{20 AU}}\sim 0.3\mbox{ g cm$^{-2}$}$, similar to the surface density of the Minimum Mass Solar Nebula at this distance \citep{Hayashi1981}. It seems that the cold population did not suffer a mass depletion due to scattering, nor could it have suffered order-of-magnitude mass loss due to collisional evolution \citep[see discussion by][]{Morbidelli2008}. As such, the primordial surface density in the cold region must be similar to what it is today, $\Sigma_{\textrm{40 AU}}\sim10^{-5} \mbox{ g cm$^{-2}$}$. Thus, the growth time ratio of the hot and cold populations is $\frac{t_\textrm{c}}{t_\textrm{h}}\sim10^{5}$.

Put simply, unless some unknown process dramatically depleted the mass of the cold objects without disturbing their primordially cold orbits, in the classical collisional planetesimal growth model, the growth time of the cold population was roughly 5 orders of magnitude longer than for the hot population. If the break diameter of the cold population is only a recent feature, and took the age of the Solar system to form, it would imply that the same break diameter in the hot population took only  $\sim50,000$~years to form, a nearly instantaneous time compared to the age of the Solar system. Alternatively, simulations of growth from km-sized planetesimals in the 20~AU region suggest that growth break-diameter sized objects require $10^6-10^7$ years to form \citep{Weidenschilling2008,Kenyon2012}. At those growth rates, cold break diameter-sized objects would require 10-100 times the age of the Solar system to form. It seems the classical collisional growth model cannot produce the cold objects in such a low mass environment. Some other mechanism is required.

\section{Conclusions}
From the literature we compiled all Kuiper belt survey data with well characterized photometry and detection efficiency and which provides some measure of the inclination and distance for each detected source. From these compiled data, we determined absolute r'-band magnitude distributions for the dynamically hot and cold Kuiper belt populations. We found that for both populations, the absolute magnitude distributions are well fit by broken power-law functions. Both populations exhibit similar break magnitudes, \Hbcold{} and  \Hbhot{} for the cold and hot populations respectively. Similarly, to the accuracy of the data, both populations exhibit identical faint-end slopes of $\alpha_\textrm{2}\sim0.2$. Unlike previous attempts to infer the slope of the size distribution of large objects from their luminosity functions, we find much steeper slopes of \alphaocold{} and \alphaohot{} for the cold and hot populations respectively. We found that the probability the cold and hot populations share the same absolute magnitude distributions is only 3\%. In addition, we find that the slope of the hot population absolute magnitude distribution becomes shallower than $\alpha_\textrm{1}\sim0.9$ for objects with absolute magnitudes $H_{\textrm{r'}}\lesssim3$.

We utilized the Kuiper-variant Kolmogorov-Smirnov statistic to test the likelihood that the H-distributions of the cold and hot KBOs share the same parent distribution as the Jupiter Trojans. When the bimodal albedo distribution of the hot population and the low albedos of the Trojans are correctly considered, there is no evidence that the Trojans and and hot KBO population exhibit different size distributions. The same cannot be said of the cold however, which exhibit a less than 1 in 1000 chance that they share the same size distribution as the Jupiter Trojans.

\acknowledgements
We thank H. Levison for his insightful comments and constructive criticism.

\bibliographystyle{apj}
\bibliography{astroelsart}

\begin{deluxetable}{llllll} 
	\tabletypesize{\small}
	\tablecaption{Best-fit Absolute Magnitude Distribution Parameters\tablenotemark{a} \label{tab:fits}}
	\tablehead{
	\colhead{Sample} & \colhead{$\alpha_\textrm{1}$ } & \colhead{$\alpha_\textrm{2}$} & \colhead{$H_\textrm{o}$ (r')} & \colhead{$H_\textrm{B}$ (r')} & \colhead{$P(L_{\textrm{ran}}>L_{\textrm{obs}})$}
	}
	\startdata
	Trojan \tablenotemark{b} & $1.0\pm0.2$ & $0.36\pm0.01$ & N/A & $8.4_{-0.1}^{+0.2}$ & 47\% \\
	Cold, $i\leq5^o$, $38\leq r \leq 48$~AU & $1.5_{-0.2}^{+0.4}$ & $0.38_{-0.09}^{+0.05}$ & $7.36_{-0.18}^{+0.04}$ & $6.9_{-0.2}^{+0.1}$ & 76\% \\ 
	Cold \tablenotemark{*} & $1.5$ & $0.38$ & $7.33$ & $6.9$ & \\
	Hot, $i\geq5^o$, $30\leq r$~AU & $0.87_{-0.2}^{+0.07}$ & $0.2_{-0.6}^{+0.1}$ & $7.6_{-0.1}^{+0.2}$ & $7.7_{-0.5}^{+1.0}$ & 40\% \\
	Hot \tablenotemark{*} & $0.83$ & $0.0$ & $7.7$ & $8.4$ & \\
     	\enddata
	\tablenotetext{*}{ - radial distribution taken from CFEPS model. We adopt the same parameter uncertainties and $P(L_{\textrm{ran}}>L_{\textrm{obs}})$ values as those evaluated when the $\Gamma_\textrm{c}$ are treated as free parameters.}
	\tablenotetext{a}{ - Uncertainties are the extrema of the 1-$\sigma$ likelihood contours (see Section 3)}
	\tablenotetext{b}{ - fit assuming all Trojans were observed at the same distance.}

\end{deluxetable}

\begin{figure}[h]
   \centering
   \plotone{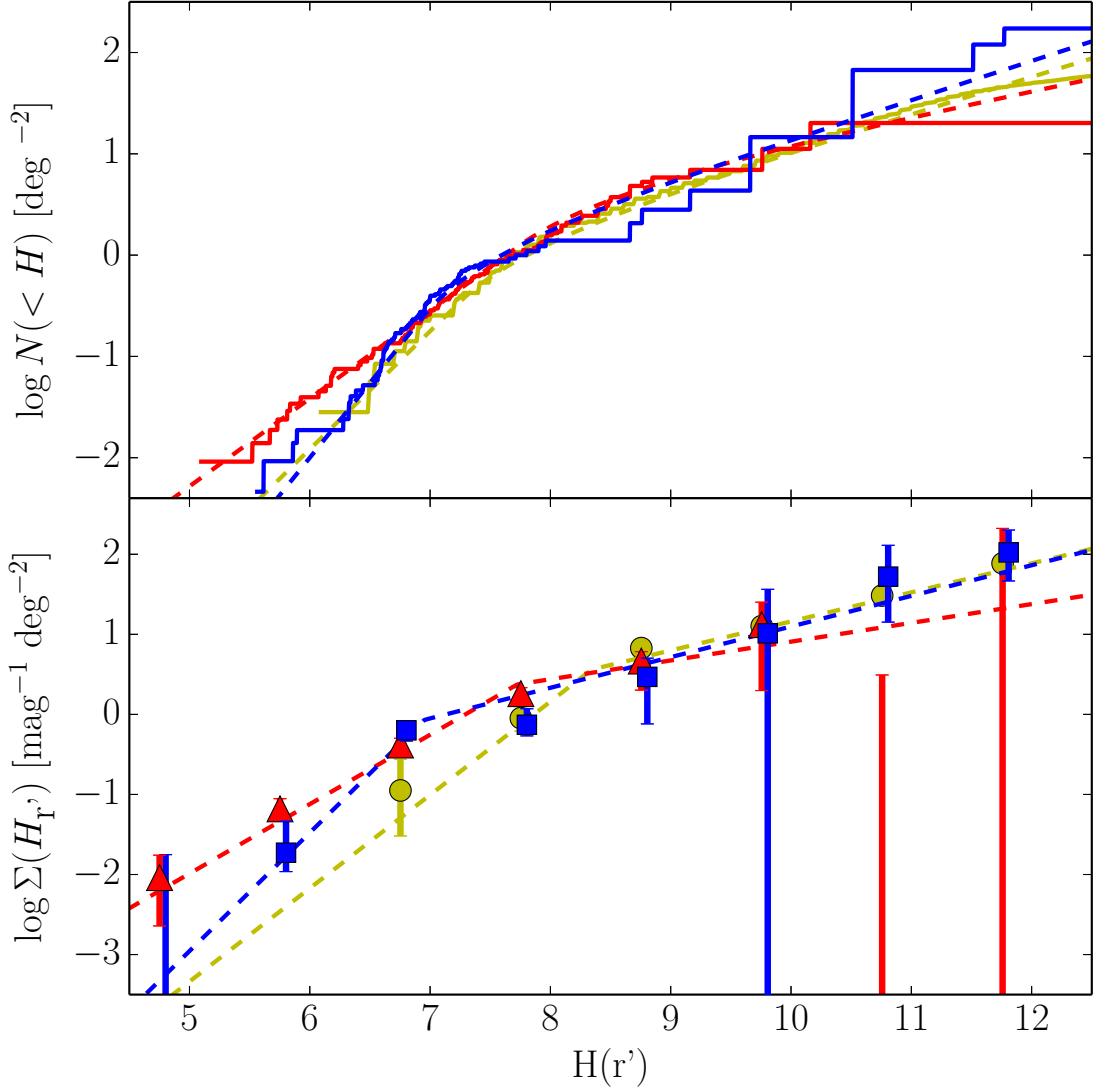} 
   \figcaption{Cumulative (top) and differential (bottom) histograms of the observed absolute magnitude distributions evaluated with use of the synthetic CFEPS radial distribution. The solid lines and points present the observed distributions. Points and lines are colour coded according to the populations they show - red triangles: \hot{} population,  blue squares: \cold{} population, yellow circles: Jupiter Trojans.. {\it Cold} population has been shifted by 0.05 magnitudes for clarity. Errorbars are the 1-$\sigma$ extents from the Monte-Carlo calculations (see Section~\ref{sec:debiasing}) and the Poissonian 1-$\sigma$ intervals added in quadrature. 2-$\sigma$ Poissonian upper limits are shown where no objects have been detected. The dashed lines represent the best-fits to the distributions (see Section~\ref{sec:fitting}).  \label{fig:hPlot_CFEPS}}
\end{figure}

\begin{figure}[h]
   \centering
   \plotone{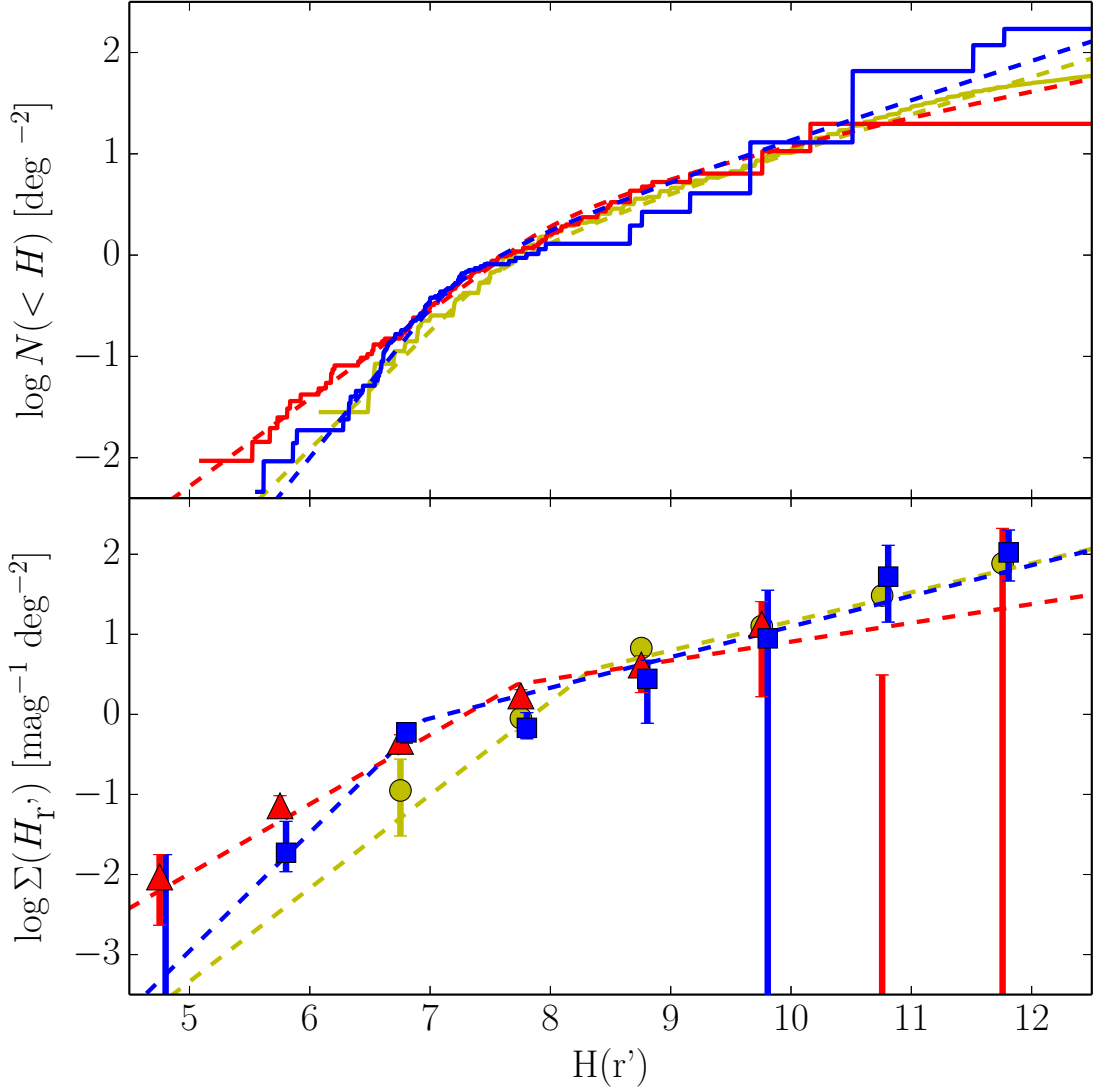} 
   \figcaption{Cumulative (top) and differential (bottom) histograms of the observed absolute magnitude distribution with the radial distribution fit from the observations.The solid lines and points present the observed distributions. Points and lines are colour coded according to the populations they show - red triangles: \hot{} population,  blue squares: \cold{} population, yellow circles: Jupiter Trojans. {\it Cold} population has been shifted by 0.05 magnitudes for clarity. Errorbars are the 1-$\sigma$ extents from the Monte-Carlo calculations (see Section~\ref{sec:debiasing}) and the Poissonian 1-$\sigma$ intervals added in quadrature. 2-$\sigma$ Poissonian upper limits are shown where no objects have been detected. The dashed lines represent the best-fits to the distributions (see Section~\ref{sec:fitting}).   \label{fig:hPlot}}
\end{figure}

\begin{figure}[h]
   \centering
   \plotone{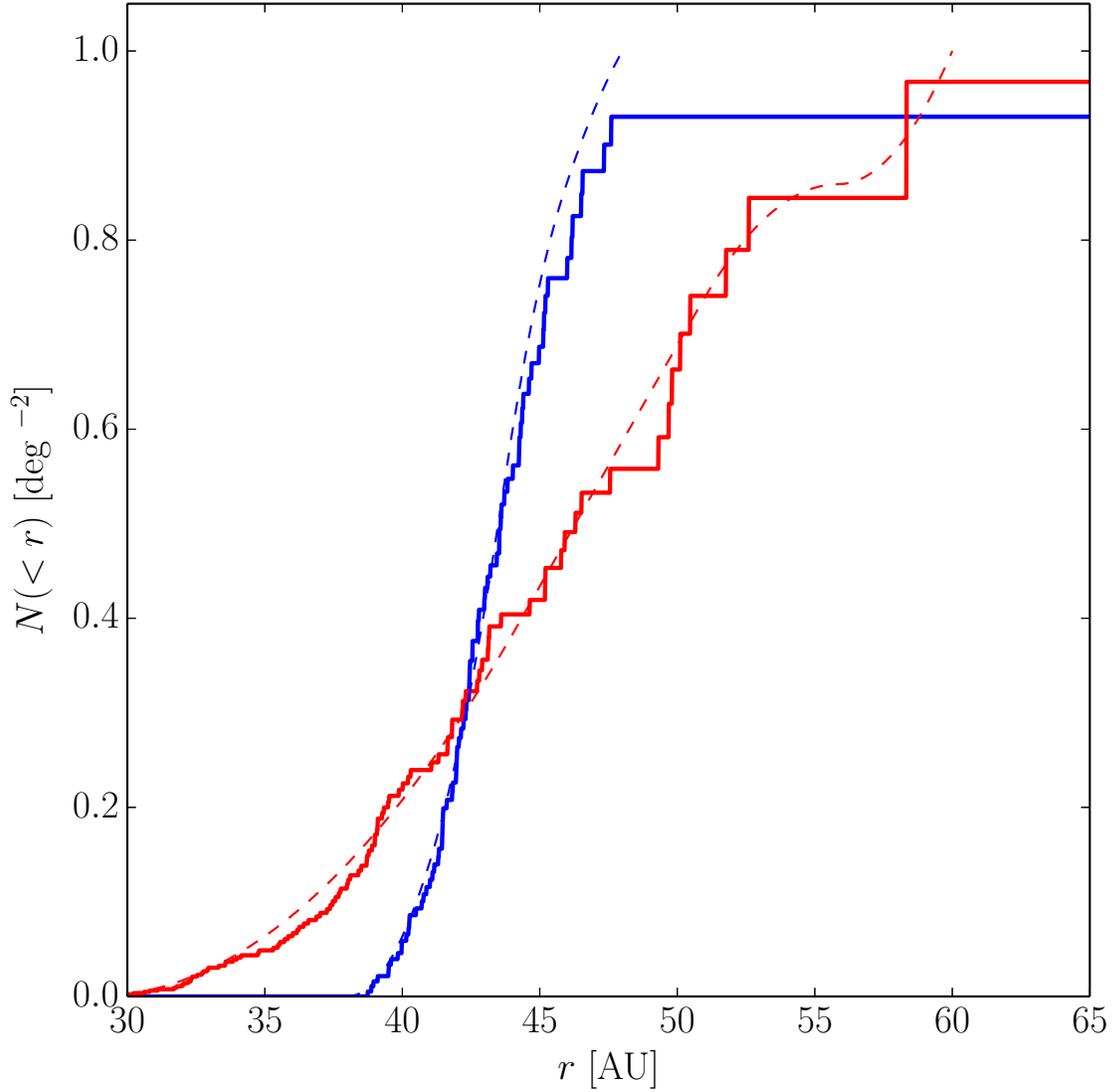} 
   \figcaption{Cumulative histograms of the observed radial distributions of the \cold{} (blue) and \hot{} samples. The solid and dashed lines present the observed and best-fit radial distributions found when the fits made use of the observed objects distances and the radial distribution parameters were free parameters in the fits.  \label{fig:rPlot}}
\end{figure}

\begin{figure}[h]
   \centering
   \plotone{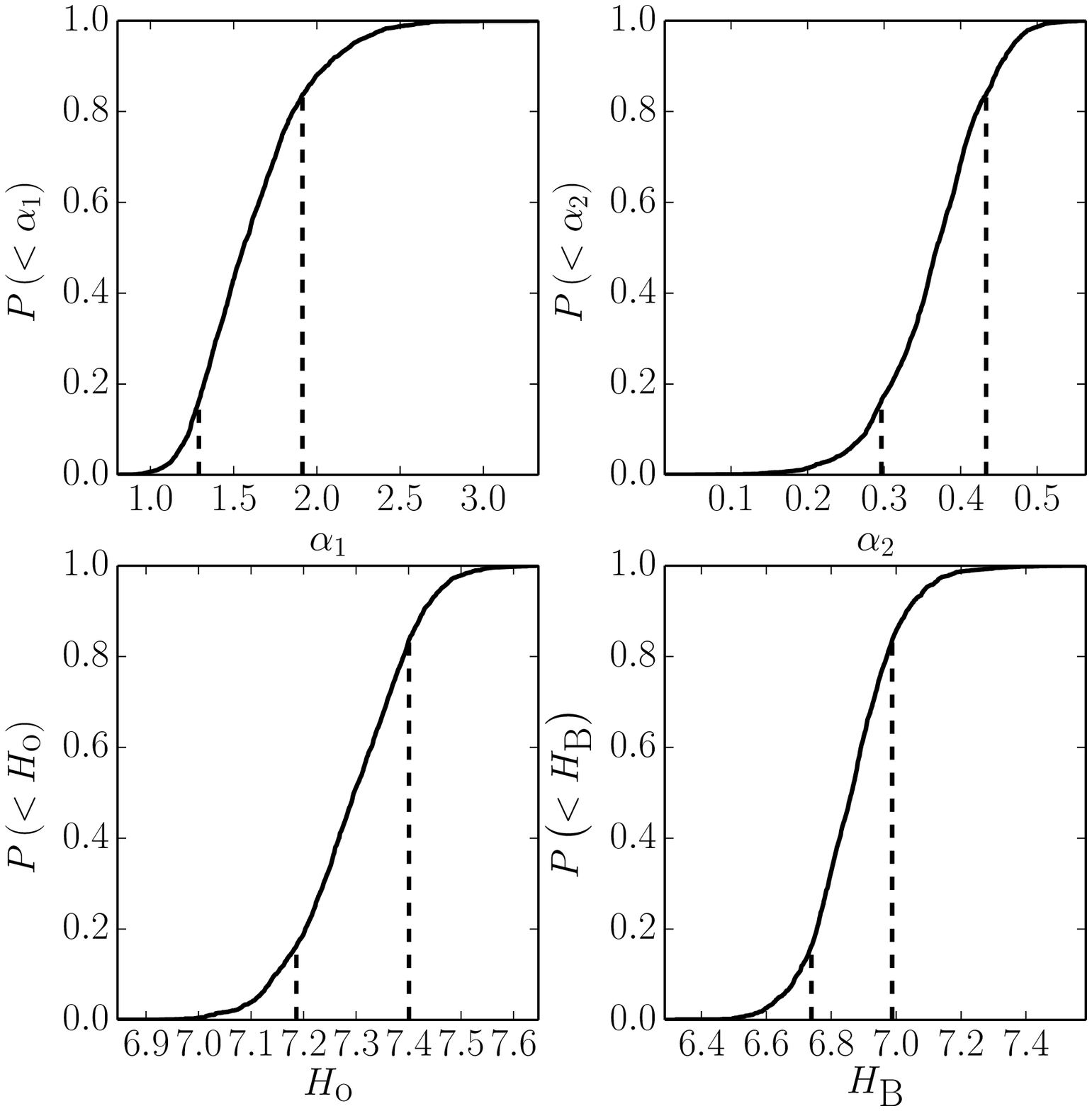} 
   \figcaption{Marginalized cumulative posterior likelihood distributions of the absolute magnitude distribution parameters for the \cold{} population. The dashed lines mark the lower and upper $1-\sigma$ bounds. \label{fig:intervalsCold}}
\end{figure}

\begin{figure}[h]
   \centering
   \plotone{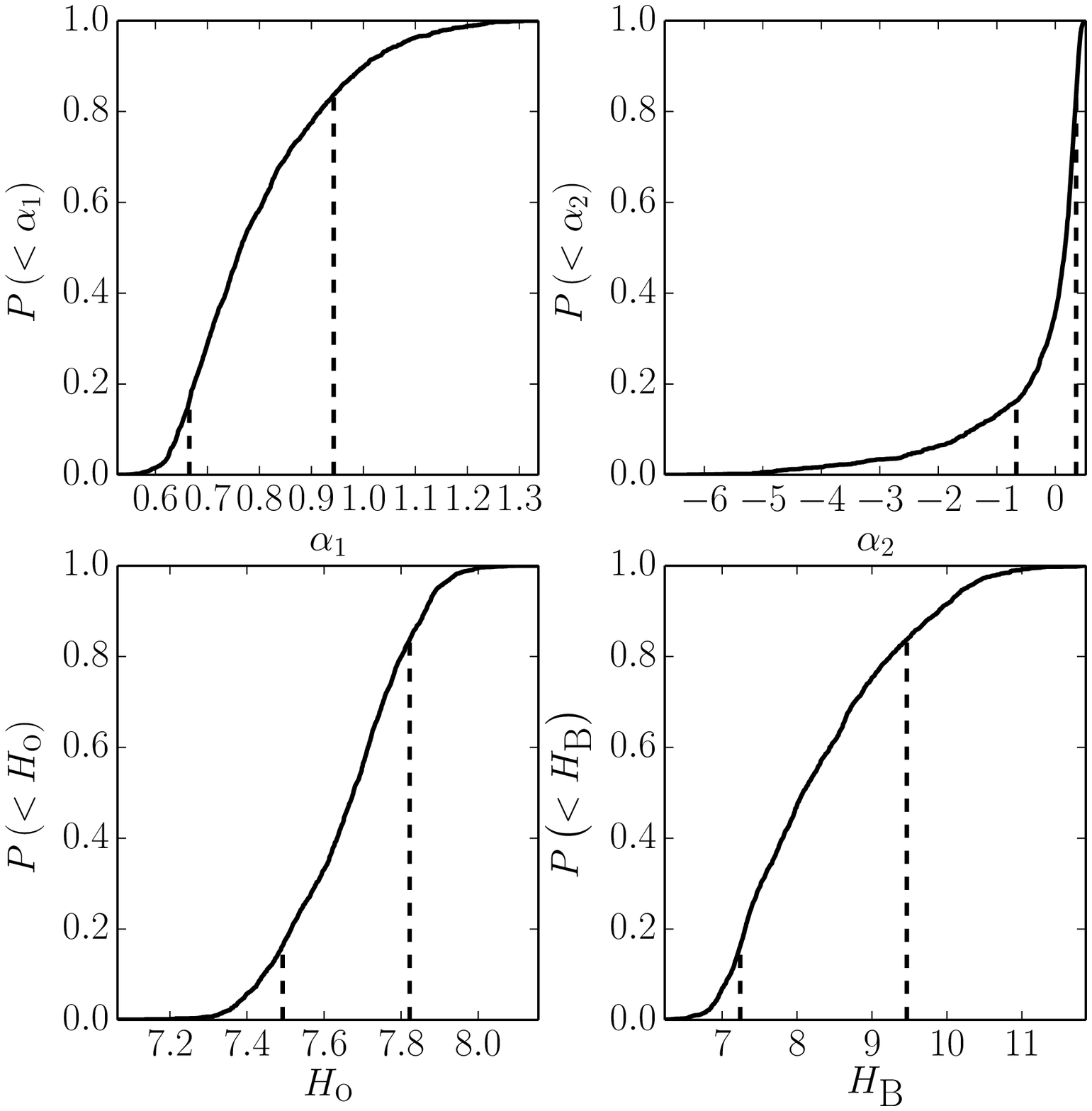} 
   \figcaption{Marginalized cumulative posterior likelihood distributions of the absolute magnitude distribution parameters for the \hot{} population. The dashed lines mark the lower and upper $1-\sigma$ bounds.  \label{fig:intervalsHot}}
\end{figure}

\begin{figure}[h]
   \centering
   \plotone{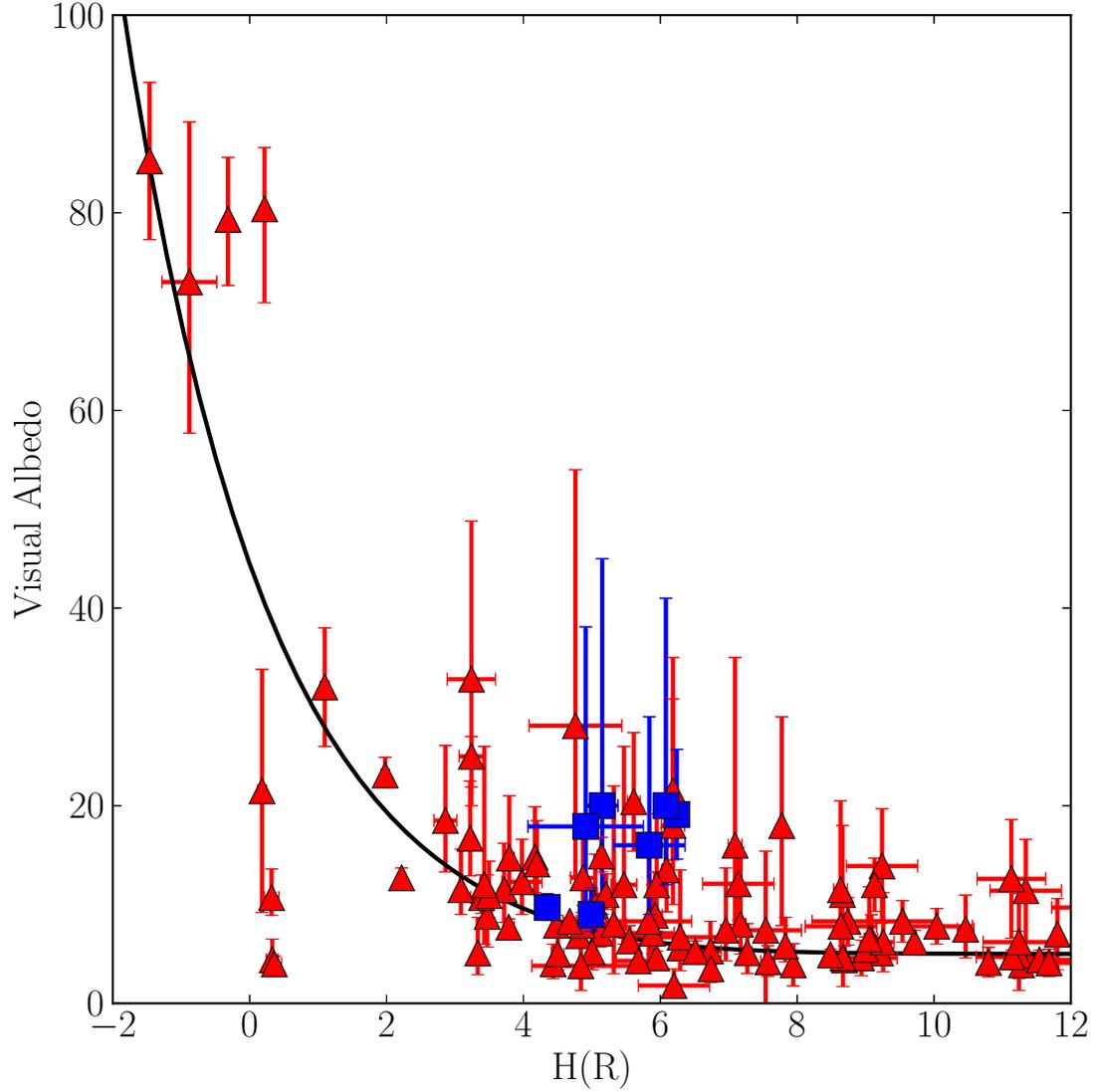} 
   \figcaption{Visual albedo versus absolute R-band magnitude. Blue squares denote the cold objects and red triangles denote hot objects. The black line represents the scaling of absolute magnitude with albedo, if albedo is given by $\rho=\left(\frac{D}{250\mbox{ km}}\right)^2 + 6\%$ which is an adequate representation of the trend albedo with absolute magnitude of the \hot{} population. \label{fig:alb_vs_H}}
\end{figure}

\begin{figure}[h]
   \centering
   \plotone{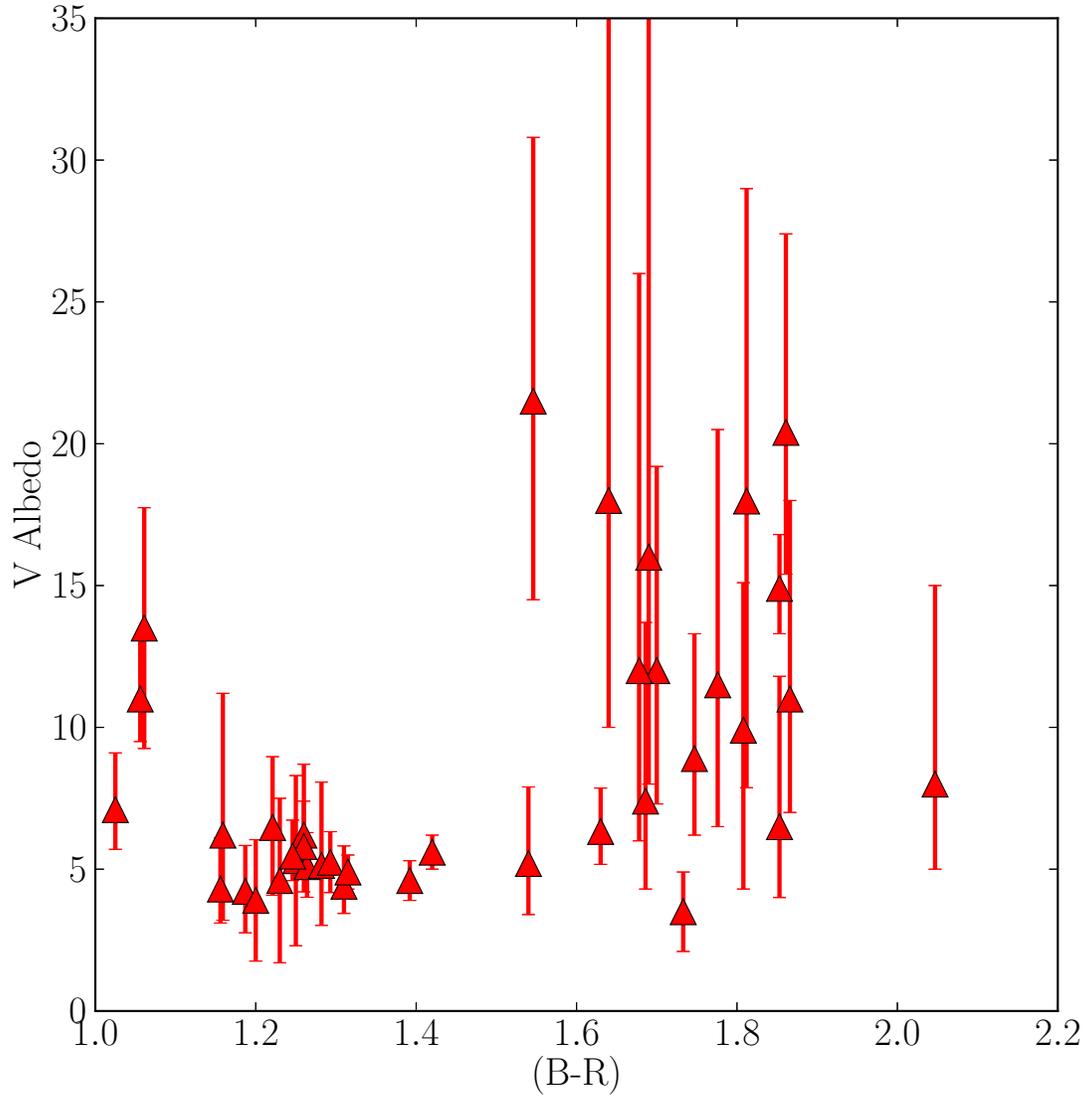} 
   \figcaption{Visual albedo versus (B-R) colour of \hot{} KBOs with $5\leq H \leq10$.   \label{fig:alb_vs_colour}}
\end{figure}

\begin{figure}[h]
   \centering
   \plotone{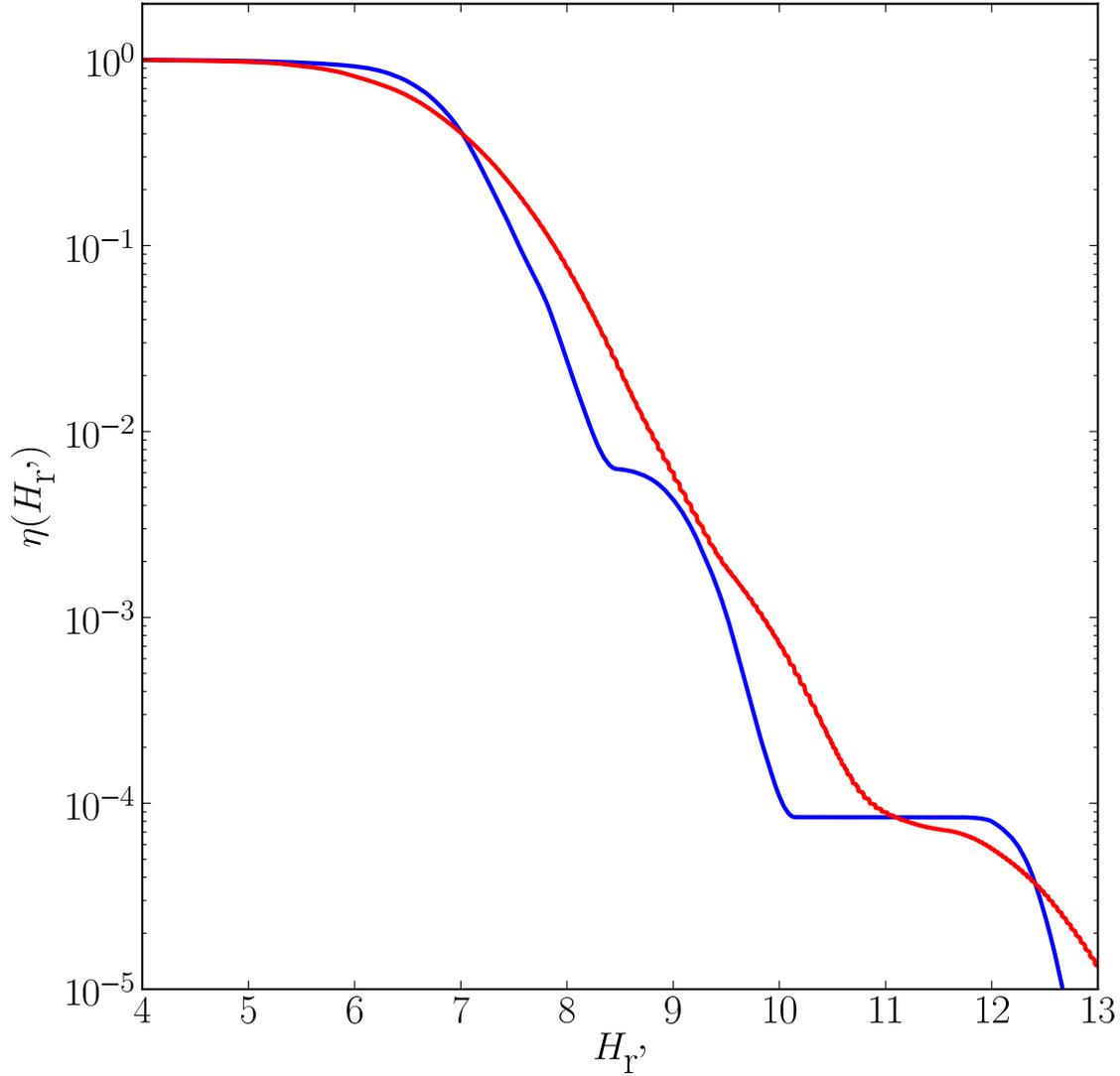} 
   \figcaption{Net effective absolute magnitude detection efficiency - found by Equation~\ref{eq:contH} - for the cold and hot KBO populations, presented as blue and red curves respectively. The net efficiency was normalized to 1 for clarity. While the net apparent magnitude efficiency $\eta(m)$ is the same for all samples, the effective absolute magnitude efficiencies, for the cold and hot populations are different as a result of their different radial distributions. \label{fig:Heff}}
\end{figure}

\begin{figure}[h]
   \centering
   \plotone{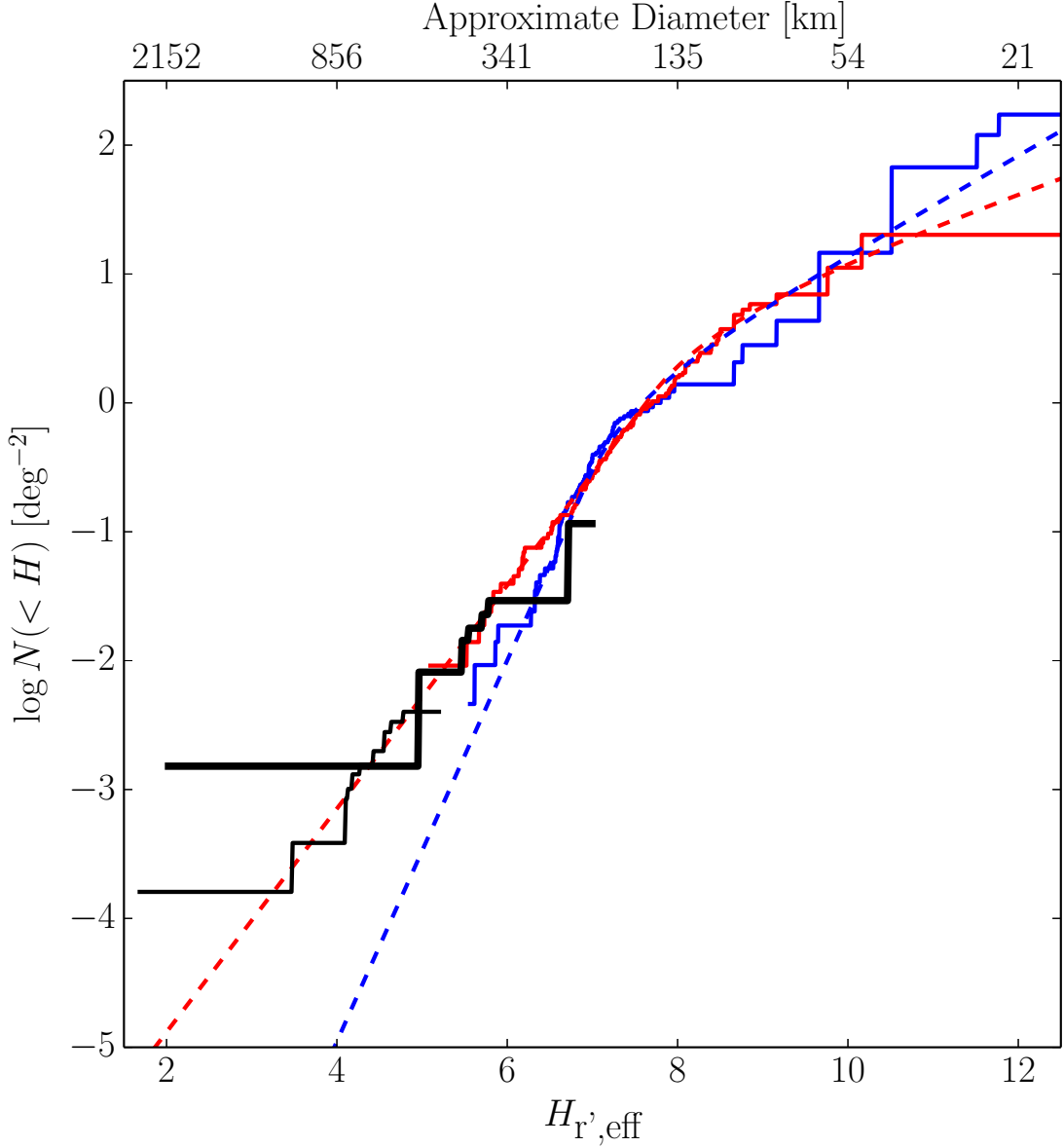} 
   \figcaption{The effective cumulative ecliptic absolute magnitude distribution of the hot population estimated from the surveys presented by \citet{Sheppard2011} (black thick) and \citet{Rabinowitz2012} (black thin) along with the ecliptic survey data of the cold (blue) and hot (red) populations. Approximate object diameters assuming 6\% albedos are presented. The black lines represent the absolute magnitudes that would be observed if the objects had albedos of 6\%. The solid lines represent the observed distributions corrected to a common albedo of 6\% for all objects, while the dashed lines represent the best-fit absolute magnitude distributions. The black lines are renormalized by small values to approximately account for the decrease in off-ecliptic sky density compared to that on the ecliptic. \label{fig:hPlot_wSheppard}}
\end{figure}

\begin{figure}[h]
   \centering
   \plotone{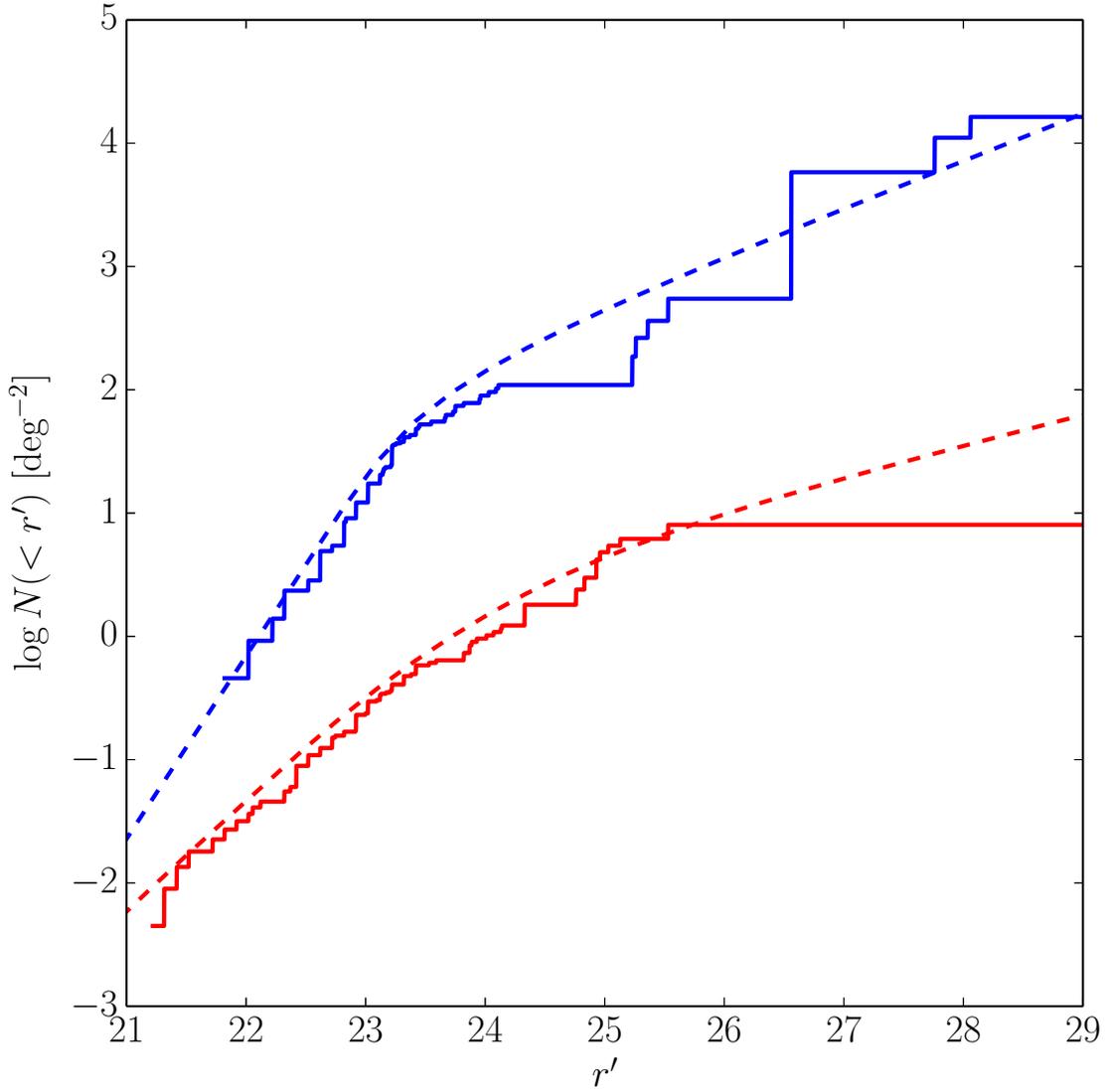} 
   \figcaption{The cumulative ecliptic luminosity functions of objects in \cold{} (blue) and \hot{} (red) samples. The solid lines represent the observed distributions while the dotted lines represent the luminosity functions determined from the best-fit H and radial distributions (see Section~\ref{sec:fitresults}). The \cold{} sample has been adjusted upwards 2 magnitudes for clarity. The estimated luminosity functions present adequate fits of the observations. \label{fig:LF}}
\end{figure}

\end{document}